\documentclass{article}
\usepackage{graphicx} 
\usepackage{epstopdf}
\usepackage{abstract}
\usepackage{appendix}
\usepackage[backref]{hyperref} 
\hypersetup{
hidelinks,
colorlinks=true,
linkcolor=black,
citecolor=black
}
\usepackage{amsmath}
\usepackage{amssymb}
\usepackage[numbers,sort&compress]{natbib}
\usepackage{enumerate}
\usepackage[a4paper, total={6in, 8in}]{geometry}
\numberwithin{equation}{section}
\usepackage{booktabs}

\usepackage{cases}
\usepackage{indentfirst}
\usepackage{algorithmic}
\usepackage[linesnumbered, ruled]{algorithm2e}
\usepackage{ntheorem}

\SetKwRepeat{Do}{do}{while}
\newtheorem{thm}{\textit{Theorem}}[section]
\newtheorem{pro}{\textit{Proposition}}[section]
\newtheorem*{myproof}{\textit{Proof}}
\newtheorem{mydef}{\textit{Definition}}[section]
\newtheorem{cor}{\textit{Corollary}}[section]
\newtheorem{lem}{\textit{Lemma}}[section]
\RequirePackage{amssymb}

\title{\textbf{Alternating Direction Method of Multipliers Based on $\ell_{2,0}$-norm for Multiple Measurement Vector Problem}}

\author{Zekun Liu, Siwei Yu\thanks{Corresponding author: siweiyu@hit.edu.cn}\\School of Mathematics, Harbin Institute of Technology 
}

\date{}

\begin{document}

\maketitle

\begin{abstract} 
The multiple measurement vector (MMV) problem is an extension of the single measurement vector (SMV) problem, and it has many applications. Nowadays, most studies of the MMV problem are based on the $\ell_{2,1}$-norm relaxation, which will fail in recovery under some adverse conditions. We propose an alternating direction method of multipliers (ADMM)-based optimization algorithm to achieve a larger undersampling rate for the MMV problem. The key innovation is the introduction of an $\ell_{2,0}$-norm sparsity constraint to describe the joint-sparsity of the MMV problem; this differs from the $\ell_{2,1}$-norm constraint that has been widely used in previous studies. To illustrate the advantages of the $\ell_{2,0}$-norm, we first prove the equivalence of the sparsity of the row support set of a matrix and its $\ell_{2,0}$-norm. Then, the MMV problem based on the $\ell_{2,0}$-norm is proposed.
Next, we give our algorithm called MMV-ADMM-$\ell_{2,0}$ by applying ADMM to the reformulated problem.
Moreover, based on the Kurdyka-Lojasiewicz property of objective functions, we prove that the iteration generated by the proposed algorithm globally converges to the optimal solution of the MMV problem. Finally, the performance of the proposed algorithm and comparisons with other algorithms under different conditions are studied with simulated examples. The results show that the proposed algorithm can solve a larger range of MMV problems even under adverse conditions.
\end{abstract}

\section{Introduction}\label{sec:1}

The multiple measurement vector (MMV) problem \cite{1} has received considerable attention in the field of compressed sensing \cite{11,12}. In the single measurement vector (SMV) problem, the purpose is to recover a vector $x\in \mathbb{R}^{n}$ from $y=Ax$, where $A\in \mathbb{R}^{m\times n}(m<n)$ and $y\in \mathbb{R}^{m}$ are given. Because $m<n$, the problem is underdetermined. With the prior information that $x$ is sparse enough \cite{13}, we can solve
\begin{equation}\label{equ:1.1}
\begin{aligned}
        &\min_{x\in \mathbb{R}^{n}}\left \| x \right \|_{0}\\
        &s.t.\enspace y=Ax,
\end{aligned}
\end{equation}
to obtain the unique solution, where the $\ell_{0}$-norm of a vector is the number of its nonzero elements. However, problem (\ref{equ:1.1}) is NP-hard and is difficult to solve directly. A common approach to overcome this drawback is to relax the $\ell_{0}$-norm to the $\ell_{1}$-norm; this is known as the basis pursuit \cite{14}:
\begin{equation}\label{equ:1.2}
\begin{aligned}
        &\min_{x\in \mathbb{R}^{n}}\left \| x \right \|_{1}\\
        &s.t.\enspace y=Ax.
\end{aligned}
\end{equation}

Problem (\ref{equ:1.2}) is convex, implying that it can be solved much more efficiently. By solving problem (\ref{equ:1.2}) we can obtain the same solution as that for problem (\ref{equ:1.1}) under suitable conditions \cite{12}.

The MMV problem is an extension of the SMV problem. It describes the case in which multiple signals are jointly measured by the same sensing matrix from signals that are jointly sparse, that is
\begin{displaymath}
    y^{j}=Ax^{j}, j=1,2,\cdots,J,
\end{displaymath}
where all of $x^{j}$ have the same locations of nonzero elements. The MMV problem occurs in many applications including hyperspectral imagery \cite{15}, computation of sparse solutions to linear inverse problems \cite{16}, neuromagnetic imaging \cite{17}, source localization \cite{18}, and equalization of sparse communication channels \cite{19}. Given $Y=(y^{1},y^{2},\cdots,y^{J})\in \mathbb{R}^{m\times J}$, the MMV problem is to recover $X=(x^{1},x^{2},\cdots,x^{J})$ from $Y=AX$ where only a few rows of $X$ have nonzero elements. For this purpose, the most widely used approach is $\ell_{2,1}$-norm minimization \cite{16,7,20}:
\begin{equation}\label{equ:1.3}
\begin{aligned}
        &\min_{X\in \mathbb{R}^{n\times J}}\left \| X \right \|_{2,1}\\
        &\quad s.t.\enspace Y=AX,
\end{aligned}
\end{equation}
and the definition of the generalized $\ell_{p,q}$-norm of a matrix is
\begin{displaymath}
    \left \| X \right \|_{p,q}=(\sum_{i=1}^{n}\left \|X_{i} \right \|_{p}^{q})^{\frac{1}{q}},
\end{displaymath}
where $X_{i}$ is the $i$th row of $X$. Owing to the benefit of the additional joint-sparsity property, the recovery performance in the accuracy and total time spent for the MMV problem are better than those of the SMV problem \cite{7,20}.

In recent decades, several recovery algorithms for the MMV problem have been proposed \cite{7,16,21,28}. \cite{21} employed the greedy pursuit strategy to recover the signals because the joint-sparsity recovery problem is NP-hard. As a similar convex relaxation technique for the SMV problem (\ref{equ:1.2}), \cite{16,7,20} used $\ell_{2,1}$-norm minimization to solve the MMV problem. For hyperspectral imagery applications, \cite{15} applied the alternating direction method of multipliers (ADMM) to solve problem (\ref{equ:1.3}). Most theoretical results on the relaxation of the SMV problem can be extended to the MMV problem \cite{7}. However, as (\ref{equ:1.2}) is a convex relaxation of (\ref{equ:1.1}) that yields the same solution only under suitable conditions \cite{12}, (\ref{equ:1.3}) is only a convex relaxation of the $\ell_{2,0}$-norm minimization problem for MMV. (\ref{equ:1.3}) cannot provide an accurate solution under certain unfavorable conditions, whereas the $\ell_{2,0}$-norm minimization problem can. Therefore, some drawbacks limit the use of previous algorithms.

In this study, instead of considering the widely used convex relaxation $\ell_{2,1}$-norm minimization problem, we directly study the original $\ell_{2,0}$-norm minimization in the MMV problem. We first note the equivalence of the joint-sparsity property and the $\ell_{2,0}$-norm in the MMV problem. Then, we reformulate the MMV problem via the sparsity constraint in \cite{7}. Next, we propose our algorithm called MMV-ADMM-$\ell_{2,0}$ by applying ADMM \cite{8} to the reformulated problem. Theoretical analysis shows that the proposed algorithm is globally convergent to the unique solution of the MMV problem. Compared with existing algorithms, the proposed algorithm can solve the MMV problem when the signals are not very sparse or the number of sensors is small, and it can achieve a larger undersampling rate.

The rest of this paper is organized as follows: In Section \ref{sec:2}, we present some definitions used for the subsequent analysis. In Section \ref{sec:3}, we propose the optimization problem by reformulating the MMV problem. In Section \ref{sec:4}, we present the proposed MMV-ADMM-$\ell_{2,0}$ algorithm and discuss the subproblems in detail. In Section \ref{sec:5}, we present the convergence results obtained with the proposed algorithm. In Section \ref{sec:6}, we describe numerical experiments performed to verify the validity of the proposed algorithm and compare it with other algorithms. In Section \ref{sec:7}, we present the conclusions of this study.

\section{Preliminaries}\label{sec:2}

We list some of the notations and definitions used for the subsequent analysis. For any vector $x=
\left ( x_{1},x_{2},\cdots,x_{N} \right )^{T}\in \mathbb{R}^{N}$, its sparse support set is defined as
\begin{displaymath}
Supp(x)=\left \{ i|x_{i} \ne 0 \right \} \subseteq \left \{1,2,\cdots,N  \right \}.
\end{displaymath}

Note that the assumption of the MMV problem is that the set of all vectors $\left \{ x^{j}\right \}_{j=1}^{J} \in \mathbb{R}^{N}$ has the same sparse support set, implying that
\begin{displaymath}
    Supp(x^{1})=Supp(x^{2})=\cdots=Supp(x^{J}).
\end{displaymath}

\begin{mydef}\label{def:2.1}
For a set of vectors $\left \{ x^{j}\right \}_{j=1}^{J} \in \mathbb{R}^{N}$, the common sparse support set is defined as
\begin{displaymath}
    Supp(\left \{ x^{j}\right \}_{j=1}^{J})=\bigcup_{j=1}^{J} Supp(x^{j}).
\end{displaymath}
\end{mydef}

In fact, although it is almost impossible to satisfy the above assumption with real datasets, the MMV problem works well when real data have sufficient joint-sparsity \cite{7}.

The following few definitions are used for the subsequent analysis.

\begin{mydef}[\cite{4}]\label{def:2.2}
Let $f$ be a generalized real function.
\begin{enumerate}[(i)]
\item For a nonempty set $\mathcal{X}$, we call $f$ proper to $\mathcal{X}$ if there exist $x\in \mathcal{X}$ such that $f(x)<+\infty$ and for any $x\in \mathcal{X}, f(x)>-\infty$.
\item $f$ is a lower semicontinuous function if for any $x\in \mathbb{R}^{m\times n}, \liminf_{y\to x}f(y)\ge f(x)$.
\item $f$ is a closed function if its epigraph 
\begin{displaymath}
    epif=\left \{ (x,t)\in \mathbb{R}^{m\times n}\times \mathbb{R}|f(x)\le t\right \}
\end{displaymath}
is a closed set.
\end{enumerate}
\end{mydef}

For a generalized real function $f$, the lower semicontinuity of $f$ is equivalent to it being a closed function. 

\begin{mydef}[\cite{3}]\label{def:2.3}
For a proper lower semicontinuous function $f$, its critical point is one that satisfies $0\in \partial f(x)$.
\end{mydef}

In keeping with the first-order optimal condition, if $x$ is a minimizer of $f$, it must be a critical point of $f$ first.

\begin{mydef}[\cite{5}]\label{def:2.4}
$S \subseteq \mathbb{R}^{m\times n}$ is a semi-algebraic set if there exist polynomial functions $f_{ij},g_{ij}:\mathbb{R}^{m\times n}\to \mathbb{R}$ with $1\le i\le p,1\le j\le q$ such that 
\begin{displaymath}
    S=\bigcup_{i=1}^{p}\bigcap_{j=1}^{q}\left \{x\in  \mathbb{R}^{m\times n}:f_{ij}(x)=0,g_{ij}(x)>0 \right \}.
\end{displaymath}
The proper function $f$ is semi-algebraic if its graph
\begin{displaymath}
    graphf=\left \{(x,t)\in \mathbb{R}^{m\times n}\times\mathbb{R}:f(x)=t \right \}
\end{displaymath}
is a semi-algebraic set of $\mathbb{R}^{m\times n}\times\mathbb{R}$.    
\end{mydef}

Semi-algebra has a few useful properties:
\begin{enumerate}[(i)]
    \item Semi-algebraic sets are closed under finite unions and Cartesian products.
    \item Many common functions are actually semi-algebraic, including real polynomials, compositions of semi-algebraic functions, indicator functions of semi-algebraic sets, and $\pi(A)$ where $A\subseteq \mathbb{R}^{m\times n}\times\mathbb{R}$ is a semi-algebraic set and $\pi:\mathbb{R}^{m\times n}\times\mathbb{R}\to \mathbb{R}^{m\times n}$ is the projection operator.
\end{enumerate}

The definitions of Gâteaux differentiable, subdifferential, Karush-Kuhn-Tucker (KKT) conditions, and Kurdyka-Lojasiewicz (KL) property can be referred elsewhere \cite{2,3,4,5,6}.

The following proposition states the connection between semi-algebraic functions and KL functions.

\begin{pro}[\cite{6}]\label{pro:2.1}
If the proper lower semicontinuous function $f$ is semi-algebraic, it must be a KL function.
\end{pro}

\section{Problem formulation}\label{sec:3}

First, we state a proposition that is useful to describe the MMV problem in mathematical language.

\begin{pro}\label{pro:3.1}
For a set of vectors $\left \{  s^{j}\right \}_{j=1}^{J}\in \mathbb{R}^{N}$, denote $S=(s^{1},s^{2},\cdots,s^{J})$. Then, $\left \{  s^{j}\right \}_{j=1}^{J}$ has the common sparse support set with sparsity $k \Longleftrightarrow \left \| S \right \|_{2,0}=k$.
\end{pro}

\begin{myproof}
Denote the common sparse support set of $\left \{  s^{j}\right \}_{j=1}^{J}$ as $Supp$. $S$ can be expressed by its row vector as $S=(\alpha_{1},\alpha_{2},\cdots,\alpha_{N})^{T}$. Then, we can conclude that
 \begin{align*}
 &\left \| S \right \|_{2,0}=\left \| (\left \| \alpha_{1} \right \|_{2},\left \| \alpha_{2} \right \|_{2},\cdots,\left \| \alpha_{N} \right \|_{2} )^{T} \right \|_{0}=k \\
\Longleftrightarrow & \begin{cases}
 \left \| \alpha_{j} \right \|_{2}\ne 0 \quad \text{for} \enspace  j\in Supp\\
\left \| \alpha_{j} \right \|_{2}= 0 \quad \text{for} \enspace  j\notin Supp
\end{cases} \quad and \enspace \left | Supp \right |=k \quad \text{(definition of $\ell_{0}$-norm)} \\
 \\
\Longleftrightarrow & \begin{cases}
 \alpha_{j}\ne 0 \quad \text{for} \enspace  j\in Supp\\
\alpha_{j}= 0 \quad \text{for} \enspace  j\notin Supp
\end{cases} \quad \text{(positivity of $\ell_{2}$-norm)} \\
 \\
\Longleftrightarrow & Supp \enspace \text{is actually the common sparse support set with}\enspace \left | Supp \right |=k.
\end{align*}
\noindent This completes the whole proof.   \hfill $\square$
\end{myproof}

Next, we convert the MMV problem to an optimization problem by using Proposition \ref{pro:3.1}.

Assuming there are $J$ sensors, the sparse vectors after sparse representation are $s_{1},s_{2},\cdots,s_{J}$, where $s_{j}\in \mathbb{R}^{N}$ for all $j\in \left \{ 1,2,\cdots,J \right \}$; we denote $S=(s_{1},s_{2},\cdots,s_{J})\in \mathbb{R}^{N\times J}$. Assuming the sensing matrix $\Phi \in \mathbb{R}^{M\times N}(M<N)$, the measurement vectors are $y_{1},y_{2},\cdots,y_{J}$, where $y_{j}=\Phi s_{j}\in \mathbb{R}^{M}$ for all $j\in \left \{ 1,2,\cdots,J \right \}$; we denote $Y=(y_{1},y_{2},\cdots,y_{J})\in \mathbb{R}^{M\times J}$. By Proposition \ref{pro:3.1}, minimizing the joint-sparsity of $\left \{ s_{j} \right \}_{j=1}^{J}$ is equivalent to minimizing $\left \| S \right \|_{2,0}$. Therefore, the MMV problem can be described as
\begin{equation}\label{equ:3.1}
    \begin{aligned}
        &\min_{S\in \mathbb{R}^{N\times J}} \left \| S \right \|_{2,0}\\
        &\quad s.t.\quad Y=\Phi S,
    \end{aligned}
\end{equation}

The following theorem offers the prerequisite to make a further conversion of problem (\ref{equ:3.1}).

\begin{thm}[\cite{7}]\label{thm:3.1}
For $\Phi \in \mathbb{R}^{M\times N}$ and $S\in \mathbb{R}^{N\times J}$, if $Y=\Phi S$ and 
\begin{equation}\label{equ:3.2}
    \left \| S \right \|_{2,0}<\frac{Spark(\Phi)+Rank(Y)-1}{2},
\end{equation}
then problem (\ref{equ:3.1}) will only possess the unique solution $S$.
\end{thm}

Based on Theorem \ref{thm:3.1}, we do not require all vectors to share the same sparse support set in the MMV problem. Instead, we just require that $S$ satisfies (\ref{equ:3.2}).

To consider the odevity of $Spark(\Phi)+Rank(Y)$ and the measurement error between $Y$ and the real $\Phi S$, set 
\begin{equation}\label{equ:3.3}
    s=\left \lfloor \frac{Spark(\Phi)+Rank(Y)-2}{2} \right \rfloor,
\end{equation}
where $\left \lfloor \cdot \right \rfloor$ is the round down operator. Problem (\ref{equ:3.1}) can be converted to
\begin{equation}\label{equ:3.4}
    \begin{aligned}
        &\min_{S\in \mathbb{R}^{N\times J}} \left \| Y-\Phi S \right \|_{F}^{2}\\
        &\quad s.t.\quad \left \| S \right \|_{2,0}\le s.
    \end{aligned}
\end{equation}

It is a nonconvex constrained optimization problem. The indicator function
\begin{displaymath}
    \mathcal{I}_{\mathcal{M}}(X)=\begin{cases}
 0, & \text{ \textit{if} } X\in \mathcal{M} \\
 +\infty, & \text{ \textit{if} } X\notin \mathcal{M}
\end{cases}
\end{displaymath}
is introduced to the set
\begin{displaymath}
    \mathcal{M}=\left \{ X\in \mathbb{R}^{N\times J}:\left \| X \right \|_{2,0}\le s \right \}
\end{displaymath}
to move the nonconvex constraint to the objective function, and matrix $B\in \mathbb{R}^{N\times J}$ is introduced as an auxiliary variable of $S$ to reformulate problem (\ref{equ:3.4}) as
\begin{equation}\label{equ:3.5}
    \begin{aligned}
        &\min_{B,S\in \mathbb{R}^{N\times J}} \left \| Y-\Phi S \right \|_{F}^{2}+\mathcal{I}_{\mathcal{M}}(B)\\
        &\quad s.t.\quad \quad B-S=0.
    \end{aligned}
\end{equation}

Now we have obtained the final problem formulation (\ref{equ:3.5}) to describe the MMV problem.

\section{Algorithm}\label{sec:4}

Problem (\ref{equ:3.5}) is a two block nonconvex optimization problem with a linear equality constraint; it can be solved using ADMM \cite{8}.

The augmented Lagrangian function associated with problem (\ref{equ:3.5}) is defined as
\begin{equation}\label{equ:4.1}
    \mathcal{L}_{\rho}(B,S,L)= \left \| Y-\Phi S \right \|_{F}^{2}+\mathcal{I}_{\mathcal{M}}(B)+\left \langle L,B-S  \right \rangle +\frac{\rho}{2}\left \| B-S \right \|_{F}^{2},
\end{equation}
where $L\in \mathbb{R}^{N\times J}$ is the Lagrangian multiplier of the equation constraint, $\left \langle \cdot,\cdot  \right \rangle$ is the inner product of matrices, and $\rho>0$ is the penalty parameter.

By using the ADMM scheme, we can obtain an approximate solution of problem (\ref{equ:3.5}) by minimizing the variables one by one with others fixed, as follows:
\begin{equation}\label{equ:4.2}
    \begin{aligned}
        \left\{\begin{matrix}
 B^{k+1}=&\mathop{\arg\min}\limits_{B\in \mathbb{R}^{N\times J}}\mathcal{L}_{\rho}(B,S^{k},L^{k}),\\
 S^{k+1}=&\mathop{\arg\min}\limits_{S\in \mathbb{R}^{N\times J}}\mathcal{L}_{\rho}(B^{k+1},S,L^{k}),\\
L^{k+1}=&L^{k}+\rho (B^{k+1}-S^{k+1}).
\end{matrix}\right.
    \end{aligned}
\end{equation}

Next, the subproblems in Equation (\ref{equ:4.2}) are solved one by one.

\subsection{Update B}\label{sub:4.1}

Fix $S$ and $L$,
\begin{equation}\label{equ:4.3}
    \begin{split}
        B^{k+1}&=\mathop{\arg\min}\limits_{B\in \mathbb{R}^{N\times J}}\mathcal{L}_{\rho}(B,S^{k},L^{k})\\
        &=\mathop{\arg\min}\limits_{B\in \mathbb{R}^{N\times J}}\left \{  \mathcal{I}_{\mathcal{M}}(B)+\left \langle L^{k},B-S^{k}  \right \rangle +\frac{\rho}{2}\left \| B-S^{k} \right \|_{F}^{2} \right \} \\
        &=\mathop{\arg\min}\limits_{B\in \mathcal{M}}\left \| B-S^{k}+\frac{L^{k}}{\rho} \right \|_{F}^{2}\\
        &=\mathcal{P}_{\mathcal{M} }(S^{k}-\frac{L^{k}}{\rho}),
    \end{split}
\end{equation}
where $\mathcal{P}_{\mathcal{M}}(\cdot)$ is the projection operator on set $\mathcal{M}=\left \{ X\in \mathbb{R}^{N\times J}:\left \| X \right \|_{2,0}\le s \right \}$.

Actually, $\mathcal{P}_{\mathcal{M}}(\cdot)$ is the hard thresholding operator \cite{27}. If $\left \| S^{k}-\frac{L^{k}}{\rho} \right \|_{2,0}\le s$, then $B^{k+1}=S^{k}-\frac{L^{k}}{\rho}$; otherwise, truncate $S^{k}-\frac{L^{k}}{\rho}$ with the rows whose $\ell_{2}$-norm is the top $s$ large preserved.

\subsection{Update S}\label{sub:4.2}

Fix $B$ and $L$,
\begin{equation}\label{equ:4.4}
    \begin{split}
        S^{k+1}&=\mathop{\arg\min}\limits_{S\in \mathbb{R}^{N\times J}}\mathcal{L}_{\rho}(B^{k+1},S,L^{k})\\
        &=\mathop{\arg\min}\limits_{S\in \mathbb{R}^{N\times J}}\left \{ \left \| Y-\Phi S \right \|_{F}^{2}  +\left \langle L^{k},B^{k+1}-S\right \rangle +\frac{\rho}{2}\left \| B^{k+1}-S\right \|_{F}^{2} \right \} \\
        &=\mathop{\arg\min}\limits_{S\in \mathbb{R}^{N\times J}}\left \{ \left \| Y-\Phi S \right \|_{F}^{2}+\frac{\rho}{2}\left \| B^{k+1}-S+\frac{L^{k}}{\rho} \right \|_{F}^{2} \right \},
    \end{split}
\end{equation}

Denote $f(X)=\left \| Y-\Phi X \right \|_{F}^{2}+\frac{\rho}{2}\left \| B^{k+1}-X+\frac{L^{k}}{\rho} \right \|_{F}^{2}$. The first-order optimal condition for unconstrained differentiable optimization problem gives
\begin{displaymath}
    S^{k+1}=\mathop{\arg\min}\limits_{S\in \mathbb{R}^{N\times J}}f(S) \Longleftrightarrow  \nabla f(S^{k+1})=0.
\end{displaymath}

It is easy to calculate
\begin{displaymath}
    \nabla f(S^{k+1})=(2\Phi^{T}\Phi+\rho I)S^{k+1}-2\Phi^{T}Y-\rho B^{k+1}-L^{k}.
\end{displaymath}

Therefore, by solving the linear equation $\nabla f(S^{k+1})=0$, we obtain
\begin{equation}\label{equ:4.5}
    S^{k+1}=(2\Phi^{T}\Phi+\rho I)^{-1}(2\Phi^{T}Y+\rho B^{k+1}+L^{k}).
\end{equation}

\subsection{Update L}\label{sub:4.3}

Fix $B$ and $S$. The Lagrangian multiplier $L$ can be updated as
\begin{equation}\label{equ:4.6}
    L^{k+1}=L^{k}+\rho (B^{k+1}-S^{k+1}).
\end{equation}

\subsection{Convergence criterion}\label{sub:4.4}

The convergence criterion of the proposed algorithm is given below, and the reasons will be discussed in the following section.

Assuming $\left \{ (B^{k},S^{k},L^{k}) \right \}$ is the sequence generated using the ADMM procedure (\ref{equ:4.2}), the convergence criterion is given as
\begin{equation}\label{equ:4.7}
        \begin{cases}
\left \| B^{k}-S^{k} \right \|_{F}\to 0,\\
 \left \| S^{k+1}-S^{k} \right \|_{F}\to 0, \\
\left \| L^{k} \right \|_{F}\to 0.
\end{cases}
\end{equation}

We call our algorithm for solving problem (\ref{equ:3.5}) the MMV-ADMM-$\ell_{2,0}$; it is summarized in Algorithm \ref{alg:1}. Section 
\ref{sec:6} describes the parameter initialization.

\begin{algorithm}[H]
  \KwIn{The measurement data matrix $Y$, the sensing matrix $\Phi$;}
  \KwOut{The reconstructed sparse matrix $\Hat{S}$;}
  initialization:$S^{0},L^{0},s,\rho$, and let $k=0$;
  
    \Do{not satisfy the convergence criterion}{
      Update $B$ by: $B^{k+1}=\mathcal{P}_{\mathcal{M} }(S^{k}-\frac{L^{k}}{\rho})$;
      
      Update $S$ by: $S^{k+1}=(2\Phi^{T}\Phi+\rho I)^{-1}(2\Phi^{T}Y+\rho B^{k+1}+L^{k})$;
      
      Update the Lagrangian multiplier $L$: $L^{k+1}=L^{k}+\rho (B^{k+1}-S^{k+1})$;
      
      Update $k: k=k+1$;
    }
    
    \Return $\Hat{S}=B^{k}$.
  \caption{MMV-ADMM-$\ell_{2,0}$}
  \label{alg:1}
\end{algorithm}

Notably, $2\Phi^{T}\Phi+\rho I$ is unchanged in the iteration; therefore, its inverse needs to be calculated only once. Algorithm \ref{alg:1} can be accelerated in an obvious way. When updating $S$ by (\ref{equ:4.5}), we need to calculate the inverse of an $N\times N$ matrix; the computational time required for this purpose will be high when $N$ is large. Because $M<N$ in compressed sensing, we can use the Sherman-Morrison-Woodbury (SMW) formula \cite{24,25} to simplify the calculation of the inverse. Specifically, based on the SMW-formula, we have
\begin{equation}\label{equ:4.8}
    (2\Phi^{T}\Phi+\rho I)^{-1}=\frac{I}{\rho}-\frac{2\Phi^{T}(I+\frac{2\Phi \Phi^{T}}{\rho})^{-1}\Phi}{\rho^{2}}.
\end{equation}
It converts the calculation of an $N\times N$ matrix inverse to the inverse of an $M\times M$ matrix. Because $M$ is much smaller than $N$ in compressed sensing, this will reduce the computational time. We call our algorithm MMV-ADMM-$\ell_{2,0}$-SMW when using (\ref{equ:4.8}) to update $S$.

\section{Convergence analysis}\label{sec:5}

Algorithm \ref{alg:1} is a two-block ADMM for a nonconvex problem; its convergence analysis remains an open problem. However, owing to the KL property of objective functions in problem (\ref{equ:3.5}), Algorithm \ref{alg:1} has global convergence.

First, we prove some properties of objective functions in problem (\ref{equ:3.5}).

\begin{pro}\label{pro:5.1}
The $\ell_{2,0}$-norm of a matrix is proper and lower semicontinuous.
\end{pro}

\begin{myproof}
First, we prove that the $\ell_{0}$-norm of a vector is lower semicontinuous. Denote
\begin{align*}
 f:&\mathbb{R}^{N}\to \mathbb{R}  \\
&x\mapsto \left \| x \right \|_{0}.
\end{align*}
For any $x=(x_{1},x_{2},\cdots,x_{N})^{T} \in \mathbb{R}^{N}$, assume 
\begin{displaymath}
    f(x)=\left \| x \right \|_{0}=s,
\end{displaymath}
\begin{displaymath}
    Supp(x)=\left \{ l_{1},l_{2},\cdots,l_{s} \right \} \subseteq \left \{ 1,2,\cdots,N \right \}.
\end{displaymath}
When $y^{k}=(y_{1}^{k},y_{2}^{k},\cdots,y_{N}^{k})^{T} \overset{k\to \infty}{\longrightarrow}x$, for 
\begin{displaymath}
\varepsilon=\frac{\min_{j=1,2,\cdots,s}\left | x_{l_{j}} \right |}{2}>0,
\end{displaymath}
there exist $K\in \mathbb{N}$ such that 
\begin{displaymath}
    \left | y_{i}^{k}-x_{i} \right |<\varepsilon, i=1,2,\cdots,N \text{ for all } k>K. 
\end{displaymath}
Therefore, when $k>K$,
\begin{displaymath}
    y_{l_{j}}^{k}\ne 0, j=1,2,\cdots,s,
\end{displaymath}
implying $f(y^{k})=\left \| y^{k} \right \|_{0}\ge s.$

According to the sign-preserving property of limit, we have $\liminf_{k\to \infty}f(y^{k})\ge s=f(x)$; therefore, $f$ is lower semicontinuous.

For any $S\in \mathbb{R}^{N\times J}$, $S$ can be expressed by its row vector as $S=(\alpha_{1},\alpha_{2},\cdots,\alpha_{N})^{T}$. Denote
\begin{align*}
 g:&\mathbb{R}^{N\times J}\to \mathbb{R}^{N}  \\
&S\mapsto (\left \| \alpha_{1} \right \|_{2},\left \| \alpha_{2} \right \|_{2},\cdots,\left \| \alpha_{N} \right \|_{2})^{T}.
\end{align*}
Then, the $\ell_{2,0}$-norm of a matrix 
\begin{align*}
 h:&\mathbb{R}^{N\times J}\to \mathbb{R}  \\
&S\mapsto \left \| S \right \|_{2,0}
\end{align*}
is the composition of $f$ and $g$, that is, $h=f\circ g$.

It is apparent that $g$ is continuous. Upon combining $f$, a lower semicontinuous function, we can conclude that for any $S\in \mathbb{R}^{N\times J}$, $S^{k}\to S$,
\begin{displaymath}
    \liminf_{k\to \infty}h(S^{k})=\liminf_{k\to \infty}f(g(S^{k}))\ge f(\lim_{k\to \infty}g(S^{k}))=f(g(S))=h(S).
\end{displaymath}
Therefore, $h$ is lower semicontinuous.

Apparently, the $\ell_{2,0}$-norm is proper. Therefore, it is proper and lower semicontinuous.   \hfill $\square$
\end{myproof}
 
Based on the equivalence of the semicontinuous function and closed function, Corollary \ref{cor:5.1} holds immediately.

\begin{cor}\label{cor:5.1}
The $\ell_{2,0}$-norm of a matrix is a closed function.
\end{cor}

\begin{cor}\label{cor:5.2}
The indicator function 
\begin{displaymath}
    \mathcal{I}_{\mathcal{M}}(X)=\begin{cases}
 0, & \text{ \textit{if} } X\in \mathcal{M} \\
 +\infty, & \text{ \textit{if} } X\notin \mathcal{M}
\end{cases}
\end{displaymath}
to the set
$\mathcal{M}=\left \{ X\in \mathbb{R}^{N\times J}:\left \| X \right \|_{2,0}\le s \right \}$ is also proper and lower semicontinuous.
\end{cor}

\begin{myproof}
We conclude that $\mathcal{I}_{\mathcal{M}}(X)$ is a closed function. In fact, for any sequence $(A_{k},t_{k})\in epi\mathcal{I}_{\mathcal{M}}$, we have $t_{k}\ge \mathcal{I}_{\mathcal{M}}(A_{k})$. According to Definition \ref{def:2.2}, we only need to prove if $(A_{k},t_{k})\to (A,t)$; then, $t\ge \mathcal{I}_{\mathcal{M}}(A)$.

Because $t_{k}\to t$, $k$ is large enough in only two cases:

If $\left \| A_{k} \right \|_{2,0}>s$, then $t_{k}=\mathcal{I}_{\mathcal{M}}(A_{k})=+\infty$. From $t_{k}\to t$, we have $t=+\infty$; therefore, $t\ge \mathcal{I}_{\mathcal{M}}(A)$.

If $\left \| A_{k} \right \|_{2,0}\le s$, then $t_{k}\ge \mathcal{I}_{\mathcal{M}}(A_{k})=0$. From $t_{k}\to t$, we have $t\ge 0$. Through Corollary \ref{cor:5.1}, we know that the $\ell_{2,0}$-norm of a matrix is a closed function. Therefore, 

\begin{center}
    $\left.
\begin{aligned}
&(A_{k},s)\in epi\left \| \cdot \right \|_{2,0}  \\
&(A_{k},s)\to (A,s)
\end{aligned}
\right\}
\Longrightarrow
(A,s)\in epi\left \| \cdot \right \|_{2,0},$
\end{center}
Implying that $\left \| A \right \|_{2,0}\le s$. Therefore, $t\ge0= \mathcal{I}_{\mathcal{M}}(A)$.

Further, $t\ge \mathcal{I}_{\mathcal{M}}(A)$. This proves that $\mathcal{I}_{\mathcal{M}}(X)$ is closed. It is apparent that $\mathcal{I}_{\mathcal{M}}(X)$ is proper. Therefore, $\mathcal{I}_{\mathcal{M}}(X)$ is also proper and lower semicontinuous. 
 \hfill $\square$
\end{myproof}

\begin{pro}\label{pro:5.2}
The $\ell_{2,0}$-norm of a matrix is a KL function.  
\end{pro}

\begin{myproof}
\cite{6} proved that both $\left \| \cdot \right \|_{0}$ and $\left \| \cdot \right \|_{p}$ with $p\in \mathbb{Q}_{+}$ are semi-algebraic. Denote functions $f$, $g$, and $h$ in the same way as the definitions in Proposition \ref{pro:5.1}. For any $S\in \mathbb{R}^{N\times J}$, $S$ can be expressed by its row vector as $S=(\alpha_{1},\alpha_{2},\cdots,\alpha_{N})^{T}$:
\begin{displaymath}
    g(S)=(\left \| \alpha_{1} \right \|_{2},\left \| \alpha_{2} \right \|_{2},\cdots,\left \| \alpha_{N} \right \|_{2})^{T}=\prod_{i=1}^{N} \left \{ \left \| \alpha_{i} \right \|_{2} \right \}
\end{displaymath}
is the Cartesian product of the $\ell_{2}$-norm. Because the semi-algebra is stable under a Cartesian product, $g$ is also semi-algebraic. Therefore, the composition $h=f\circ g$ is semi-algebraic.

Proposition \ref{pro:5.1} proves that $h$ is proper lower semicontinuous. Therefore, $h$ is a KL function according to Proposition \ref{pro:2.1}. \hfill $\square$
\end{myproof}

\begin{cor}\label{cor:5.3}
The indicator function 
\begin{displaymath}
    \mathcal{I}_{\mathcal{M}}(X)=\begin{cases}
 0, & \text{ \textit{if} } X\in \mathcal{M} \\
 +\infty, & \text{ \textit{if} } X\notin \mathcal{M}
\end{cases}
\end{displaymath}
to the set
$\mathcal{M}=\left \{ X\in \mathbb{R}^{N\times J}:\left \| X \right \|_{2,0}\le s \right \}$ is also a KL function.
\end{cor}

\begin{myproof}
First, we prove that $\mathcal{M}$ is semi-algebraic. Because $\left \| \cdot \right \|_{2,0}\in \mathbb{N}$, $\mathcal{M}$ can be expressed as
\begin{displaymath}
    \mathcal{M}=\left \{ X\in \mathbb{R}^{N\times J}:\left \| X \right \|_{2,0}\le s \right \}=\bigcup_{t=0}^{s}\left \{ X\in \mathbb{R}^{N\times J}:\left \| X \right \|_{2,0}=t \right \}.
\end{displaymath}

Proposition \ref{pro:5.2} proves that the $\ell_{2,0}$-norm is a semi-algebraic function. Therefore, we conclude through Definition \ref{def:2.4} that
\begin{displaymath}
    graph\left \| \cdot \right \|_{2,0}=\left \{ (X,t):\left \| X \right \|_{2,0}=t \right \}
\end{displaymath}
is a semi-algebraic set.

Denote the projection operator
\begin{align*}
 \pi:&\mathbb{R}^{N\times J}\times \mathbb{R}\to \mathbb{R}^{N\times J}  \\
&(X,t)\mapsto X.
\end{align*}
Then, 
\begin{displaymath}
    \mathcal{M}=\bigcup_{t=0}^{s}\pi(graph\left \| \cdot \right \|_{2,0}).
\end{displaymath}

Because the semi-algebra is stable under finite union and the projection operator $\pi$, $\mathcal{M}$ is semi-algebraic.

$\mathcal{I}_{\mathcal{M}}(X)$ is the indicator function to $\mathcal{M}$; it is also semi-algebraic. Corollary \ref{cor:5.2} proves that it is also proper lower semicontinuous. Therefore, by Proposition \ref{pro:2.1}, $\mathcal{I}_{\mathcal{M}}(X)$ is a KL function. \hfill $\square$  
\end{myproof}

Next, we prove that Algorithm \ref{alg:1} has global convergence based on \cite{9}.

In \cite{9} the authors discuss the following optimization problem: 
\begin{equation}\label{equ:5.1}
    \begin{aligned}
        &\min_{x,y} f(x)+g(y)\\
        &s.t.\quad Ax+y=b,
    \end{aligned}
\end{equation}
where $f$ is proper and lower semicontinuous, and $g$ is Gradient-$L$-Lipschitz continuous.

The authors make the following assumptions:
\begin{enumerate}[(i)]
    \item The two minimization subproblems in (\ref{equ:4.2}) possess solutions, and
    \item The penalty parameter $\rho>2L$,
    \item  $A^{T}A\succeq \mu I$ for some $\mu>0$.
\end{enumerate}

Set $A=-I$, $x=B$, $y=S$, $b=0$, $f(B)=\mathcal{I}_{\mathcal{M}}(B)$, and $g(S)=\left \| Y-\Phi S \right \|_{F}^{2}$ in (\ref{equ:5.1}); this gives problem (\ref{equ:3.5}). It is apparent that problem (\ref{equ:3.5}) satisfies the form of (\ref{equ:5.1}) and the above assumptions.

Next, we refer to the sufficient conditions in \cite{9} to guarantee that the sequence $\left \{ (B^{k},S^{k},L^{k}) \right \}$ generated by Algorithm \ref{alg:1} is bounded.

\begin{lem}[\cite{9}]\label{lem:5.1}
By applying the classic ADMM to (\ref{equ:5.1}), we obtain a sequence 
$\left \{w^{k}\right \}$. Suppose that 
\begin{equation}\label{equ:5.2}
    \bar{g}:=\inf_{y}\left \{ g(y)-\frac{1}{2L}\left \| \nabla g(y) \right \|^{2} \right \}>-\infty.
\end{equation}
$\left \{w^{k}\right \}$ is bounded as long as one of the following two statements holds:
\begin{enumerate}[(i)]
    \item $\liminf_{\left \|x  \right \|\to +\infty}f(x)=+\infty$, and
    \item $\inf_{x}f(x)>-\infty$ and $\liminf_{\left \|y  \right \|\to +\infty}g(y)=+\infty$.
\end{enumerate}
\end{lem}

The main result in \cite{9} is shown below.

\begin{thm}[\cite{9}]\label{thm:5.1}
Suppose that Lemma \ref{lem:5.1} holds and that $f$ and $g$ are KL functions. Then, $\left \{ w^{k} \right \}$ converges to the KKT point of (\ref{equ:5.1}).  
\end{thm}

As an application of the above theorem, Algorithm \ref{alg:1} has global convergence.

\begin{thm}\label{thm:5.2}
Algorithm \ref{alg:1} has global convergence, and the generated sequence $\left \{ (B^{k},S^{k},L^{k}) \right \}$ converges to the KKT point of problem (\ref{equ:3.5}).    
\end{thm}

\begin{myproof}
 First, we prove that $\left \{ (B^{k},S^{k},L^{k}) \right \}$ is bounded.

For $g(S)=\left \| Y-\Phi S \right \|_{F}^{2}$, we have $\nabla g(S)=2\Phi^{T}(\Phi S-Y)$ and $L=2\left \| \Phi  \right \|_{F}^{2}$.
For any $S\in \mathbb{R}^{N\times J}$,
\begin{displaymath}
    \left \| Y-\Phi S \right \|_{F}^{2}-\frac{1}{2L}\left \| 2\Phi^{T}(\Phi S-Y) \right \|_{F}^{2}\ge \left \| Y-\Phi S \right \|_{F}^{2}-\frac{4\left \| \Phi  \right \|_{F}^{2}}{2L}\left \| Y-\Phi S \right \|_{F}^{2}=0.
\end{displaymath}

Therefore, (\ref{equ:5.2}) is satisfied. Moreover, it is not hard to see that (ii) in Lemma \ref{lem:5.1} holds. Therefore, $\left \{ (B^{k},S^{k},L^{k}) \right \}$ is bounded.

Corollary \ref{cor:5.3} proves that $f(B)=\mathcal{I}_{\mathcal{M}}(B)$ is a KL function. Because $g(S)=\left \| Y-\Phi S \right \|_{F}^{2}$ is a real polynomial, it is also a KL function.

All the conditions in Theorem \ref{thm:5.1} are satisfied; therefore, $\left \{ (B^{k},S^{k},L^{k}) \right \}$ globally converges to the KKT point of (\ref{equ:3.5}).  \hfill $\square$   
\end{myproof}
 
Finally, we state the theorem about settings for the convergence criterion (\ref{equ:4.7}).

\begin{thm}\label{thm:5.3}
 Algorithm \ref{alg:1} is nearly convergent to the optimal solution of problem (\ref{equ:3.5}) if and only if its iteration sequence $\left \{ (B^{k},S^{k},L^{k}) \right \}$ satisfies the convergence criterion (\ref{equ:4.7}).   
\end{thm}

\begin{myproof}
In \cite{10}, the authors note that for the following problem
\begin{equation}\label{equ:5.3}
    \begin{aligned}
        &\min_{x} f(x)\\
        & s.t. \enspace x\in \mathcal{X},\left \| x \right \|_{0}\le k,
    \end{aligned}
\end{equation}
where $\mathcal{X}$ denotes a convex polyhedron, the optimal solution must also satisfy the KKT conditions.

It is easy to see that problem (\ref{equ:3.4}) satisfies the form of (\ref{equ:5.3}). Because problem (\ref{equ:3.5}) is equivalent to problem (\ref{equ:3.4}), the optimal point of (\ref{equ:3.5}) is also its KKT point.

Therefore, if $\left \{ (B^{k},S^{k},L^{k}) \right \}$ converges to the optimal solution $(B^{\ast},S^{\ast},L^{\ast})$, then $(B^{\ast},S^{\ast},L^{\ast})$ is a KKT point of (\ref{equ:3.5}). The Lagrangian function of (\ref{equ:3.5}) is
\begin{equation}\label{equ:5.4}
    \mathcal{L}(B,S,L)=\left \| Y-\Phi S \right \|_{F}^{2}+\mathcal{I}_{\mathcal{M}}(B)+\left \langle L,B-S \right \rangle,
\end{equation}
it satisfies the KKT conditions at $(B^{\ast},S^{\ast},L^{\ast})$. Therefore,
\begin{equation}\label{equ:5.5}
    \begin{cases}
 0 = \nabla_{S} \mathcal{L}(B^{\ast},S^{\ast},L^{\ast})=2\Phi^{T}(\Phi S^{\ast}-Y)-L^{\ast}, \\
 0 \in \partial_{B}\mathcal{L}(B^{\ast},S^{\ast},L^{\ast})=\partial_{B}\mathcal{I}_{\mathcal{M}}(B^{\ast})+L^{\ast}, \\
 0 =B^{\ast}-S^{\ast}.
    \end{cases}
\end{equation}

From $(B^{k},S^{k},L^{k})\to (B^{\ast},S^{\ast},L^{\ast})$ and the third formula of (\ref{equ:5.5}), we can conclude that $\left \| B^{k}-S^{k} \right \|_{F}\to 0$.

Denote $f$ in the same way as in the definition in Subsection \ref{sub:4.2}. When updating $S$, we have 
\begin{displaymath}
    \nabla f(S^{k+1})=(2\Phi^{T}\Phi+\rho I)S^{k+1}-2\Phi^{T}Y-\rho B^{k+1}-L^{k}=0,
\end{displaymath}

Let $k\to +\infty$. We obtain $2\Phi^{T}(\Phi S^{\ast}-Y)-L^{\ast}=0$, implying that the first formula of (\ref{equ:5.5}) is satisfied in Algorithm \ref{alg:1} without any further conditions.

From the rule of updating $B$ in (\ref{equ:4.3}), we know that $\mathcal{I}_{\mathcal{M}}(B^{k})=0$ is always satisfied during iterations in Algorithm \ref{alg:1}. According to Corollary \ref{cor:5.2}, $\mathcal{I}_{\mathcal{M}}(X)$ is proper and lower semicontinuous. Thus,
\begin{displaymath}
    0\le \mathcal{I}_{\mathcal{M}}(B^{\ast})\le \liminf_{k\to +\infty}\mathcal{I}_{\mathcal{M}}(B^{k})=0.
\end{displaymath}

Therefore, when limiting the space to all iteration points generated by Algorithm \ref{alg:1} and their accumulations, we have $\mathcal{I}_{\mathcal{M}}(X)\equiv 0$. Thus, $\partial_{B}\mathcal{I}_{\mathcal{M}}(B^{\ast})=0$. Therefore, the second formula of (\ref{equ:5.5}) holds if and only if $L^{\ast}=0$, which is equivalent to $\left \|L^{k}  \right \|_{F}\to 0$.

Thus far we have proved that the KKT conditions (\ref{equ:5.5}) are equivalent to the first and third formulas in (\ref{equ:4.7}). From $S^{k}\to S^{\ast}$, we have $\left \| S^{k+1}-S^{k} \right \|_{F}\to 0$, which is the second formula in (\ref{equ:4.7}). Therefore, the optimal point $(B^{\ast},S^{\ast},L^{\ast})$ must satisfy the convergence criterion (\ref{equ:4.7}). This completes the necessity.

For sufficiency, the convergence criterion for a generalized constrained optimization problem \cite{4} is that the distance between the adjacent iteration points tends to zero and the KKT conditions at the current iteration point are nearly satisfied. We have proved that the first and third formulas in (\ref{equ:4.7}) are equivalent to the KKT conditions. Therefore, the convergence criterion (\ref{equ:4.7}) ensures that the KKT conditions are satisfied. From combining the second formula in (\ref{equ:4.7}), which implies that the distance between the adjacent iteration points tends to zero, we know that the convergence criterion (\ref{equ:4.7}) guarantees $\left \{ (B^{k},S^{k},L^{k}) \right \}$ to converge to a KKT point of (\ref{equ:3.5}). Denote the KKT point as $(B^{\ast},S^{\ast},L^{\ast})$; we say that it is also the optimal point of (\ref{equ:3.5}).

In fact, at the KKT point $(B^{\ast},S^{\ast},L^{\ast})$, we have proved that $L^{\ast}=0$ and (\ref{equ:5.5}) are satisfied. Therefore, from the first formula of (\ref{equ:5.5}), we can conclude that 
\begin{equation}\label{equ:5.6}
    \Phi^{T}(\Phi S^{\ast}-Y)=0.
\end{equation}

The sensing matrix $\Phi\in \mathbb{R}^{M\times N}$ in the MMV problem is known to be row full rank. Thus, from (\ref{equ:5.6}), we know that $\Phi S^{\ast}-Y=0$. We have proved $\mathcal{I}_{\mathcal{M}}(B^{\ast})=0$, implying that $\left \| B^{\ast} \right \|_{2,0}\le s$. Because $B^{\ast}=S^{\ast}$, $\left \| S^{\ast} \right \|_{2,0}\le s$ too. Therefore, the KKT point $(B^{\ast},S^{\ast},L^{\ast})$ satisfies $Y=\Phi S^{\ast},\left \| S^{\ast} \right \|_{2,0}\le s$. From Theorem \ref{thm:3.1}, $S^{\ast}$ is the unique optimal point of (\ref{equ:3.1}). According to the equivalence of (\ref{equ:3.1}) and (\ref{equ:3.5}), we can conclude that $(B^{\ast},S^{\ast},L^{\ast})$ is also the unique optimal point of (\ref{equ:3.5}).

Therefore, when the convergence criterion is satisfied, Algorithm \ref{alg:1} converges to the optimal solution of (\ref{equ:3.5}).

This completes the proof. \hfill $\square$
\end{myproof}

In fact, it is not hard to find that all of the discussions above are also correct when we replace the $\ell_{2,0}$-norm with the $\ell_{p,0}$-norm with $p\in \mathbb{Q}_{+}$, implying that the proposed algorithm is also applicable to the $\ell_{p,0}$-norm case with $p\in \mathbb{Q}_{+}$. However, for the convenience of calculation, we just consider the $\ell_{2,0}$-norm case.

\section{Numerical simulations}\label{sec:6}

We designed numerical experiments to verify the validity of the proposed algorithm and theorems and then explored the performance and advantages of our MMV-ADMM-$\ell_{2,0}$ by comparing it with existing algorithms.

\subsection{Experiment setup}\label{sub:6.1}

All experiments were based on synthetic data and were performed on a PC with an Intel Core i7-8565U 1.8 GHz CPU and 20 GB RAM. The results in Subsections \ref{sub:6.4}-\ref{sub:6.7} are averaged over 100 independent experiments. Our datasets are generated as follows: sensing matrix $\Phi\in \mathbb{R}^{M\times N}$ is an $i.i.d.$ Gaussian random matrix with unit-norm columns, and the ground truth of the recovery problem $S\in  \mathbb{R}^{N\times J}$ is generated in two steps. First, $K$ rows were randomly selected as nonzero rows, with the remaining rows all being zero. Then, the elements of the selected $K$ rows were independently generated from $\mathcal{N}(0,1)$. The solution obtained from the different recovery algorithms was denoted as $\hat{S}$. The root-mean-square error (RMSE), given as 
\begin{displaymath}
    RMSE=\frac{\left \| \hat{S}-S \right \|_{F}}{\sqrt{NJ}},
\end{displaymath}        
was used to compare the recovery quality. The average running time was used to measure the computational cost. Recovery is considered successful when $RMSE<10^{-5}$.

For comparison, we test SOMP \cite{21} and MFOCUSS \cite{16}, both of which are traditional algorithms that show good performance for the MMV problem. We also include the MMV-SPG algorithm based on the $\ell_{2,1}$-norm, which is derived from a solver for large-scale sparse reconstruction called SPGL1 \cite{23}, to solve the MMV problem. Because the projection operator $\mathcal{P}_{\mathcal{M}}(\cdot)$ in (\ref{equ:4.3}) is a hard thresholding operator, we also check the SNIHT \cite{28} for the MMV problem. As the most direct comparison, MMV-ADMM-$\ell_{2,1}$ \cite{15} is the same ADMM scheme algorithm based on the $\ell_{2,1}$-norm. We introduce it in the experiments to compare the changes in performance after the $\ell_{2,1}$-norm is replaced by the $\ell_{2,0}$-norm.

The parameters of all algorithms are set as shown in Table \ref{tab:1}. For MMV-ADMM-$\ell_{2,0}$ and MMV-ADMM-$\ell_{2,1}$, $S^{0}$ and $L^{0}$ are separately initialized as an $i.i.d.$ Gaussian random matrix and a null matrix. For MMV-ADMM-$\ell_{2,0}$, if we have the prior information that the joint-sparsity of $S$ is about $K$, then we set $s$ to be slightly larger than $K$. Otherwise, we use (\ref{equ:3.3}) to initialize $s$. The spark of the matrix $\Phi$ is not easy to calculate accurately; however, an estimation given by \cite{13} is enough for the proposed algorithm.

\begin{table}[hbt!]
    \centering
    \begin{tabular}{cc}
    \hline
        Algorithm & Parameters \\\hline
        SOMP\cite{21} & Sparsity $K$ \\
        MFOCUSS\cite{16} & $\lambda=10^{-10}$ \\
        MMV-SPG\cite{23} & Not required \\
        SNIHT\cite{28} & Sparsity $K$, $MaxIter=10^{3}$  \\
        MMV-ADMM-$\ell_{2,1}$\cite{15} & $\lambda=10^{-6},\rho=10^{-5},MaxIter=10^{3}$ \\
        MMV-ADMM-$\ell_{2,0}$ & $\rho=1,MaxIter=10^{3}$\\\hline 
    \end{tabular}
    \caption{Parameter settings for the tested algorithms}
    \label{tab:1}
\end{table}

\subsection{Validity of MMV-ADMM-$\ell_{2,0}$}\label{sub:6.2}

In the first experiment, we observed the recovery quality of our MMV-ADMM-$\ell_{2,0}$. MMV-ADMM-$\ell_{2,0}$ was applied without a convergence criterion ($i.e.$, iterates to $MaxIter=10^{3}$) on datasets of size $N=500,M=150,K=50,J=10$ over 10 random repetitions. As shown in Table \ref{tab:2}, the $RMSE$ of MMV-ADMM-$\ell_{2,0}$ is close to the computer precision, and the running time is less than 1 s; this indicates that the proposed algorithm can recover the sparse signals precisely and efficiently.

\begin{table}[hbt!]
    \centering
    \begin{tabular}{ccc}
     \hline
        Data & RMSE & Time(s) \\ \hline
         1 & 8.9904e-16 & 0.6610 \\ 
         2 & 9.1155e-16 & 0.6290 \\ 
         3 & 9.1562e-16 & 0.5604 \\ 
         4 & 8.4978e-16 & 0.5965 \\ 
         5 & 9.1531e-16 & 0.6480 \\ 
         6 & 9.0638e-16 & 0.6051 \\ 
         7 & 9.9356e-16 & 0.5715 \\ 
         8 & 9.1655e-16 & 0.5679 \\ 
         9 & 7.7866e-16 & 0.6154 \\ 
         10 & 7.6170e-16 & 0.6352 \\ \hline
         Mean & 8.8484e-16 & 0.6090 \\
         Std & 6.9661e-17 & 0.0350  \\ \hline
    \end{tabular}
    \caption{The average recovery quality over 10 random experiments ($N=500,M=150,K=50,J=10$)}
    \label{tab:2}
\end{table}

\subsection{Test for convergence criterion}\label{sub:6.3}

\begin{figure}[hbt!]
	
	\begin{minipage}{0.32\linewidth}
		\vspace{3pt}
		\centerline{\includegraphics[width=\textwidth]{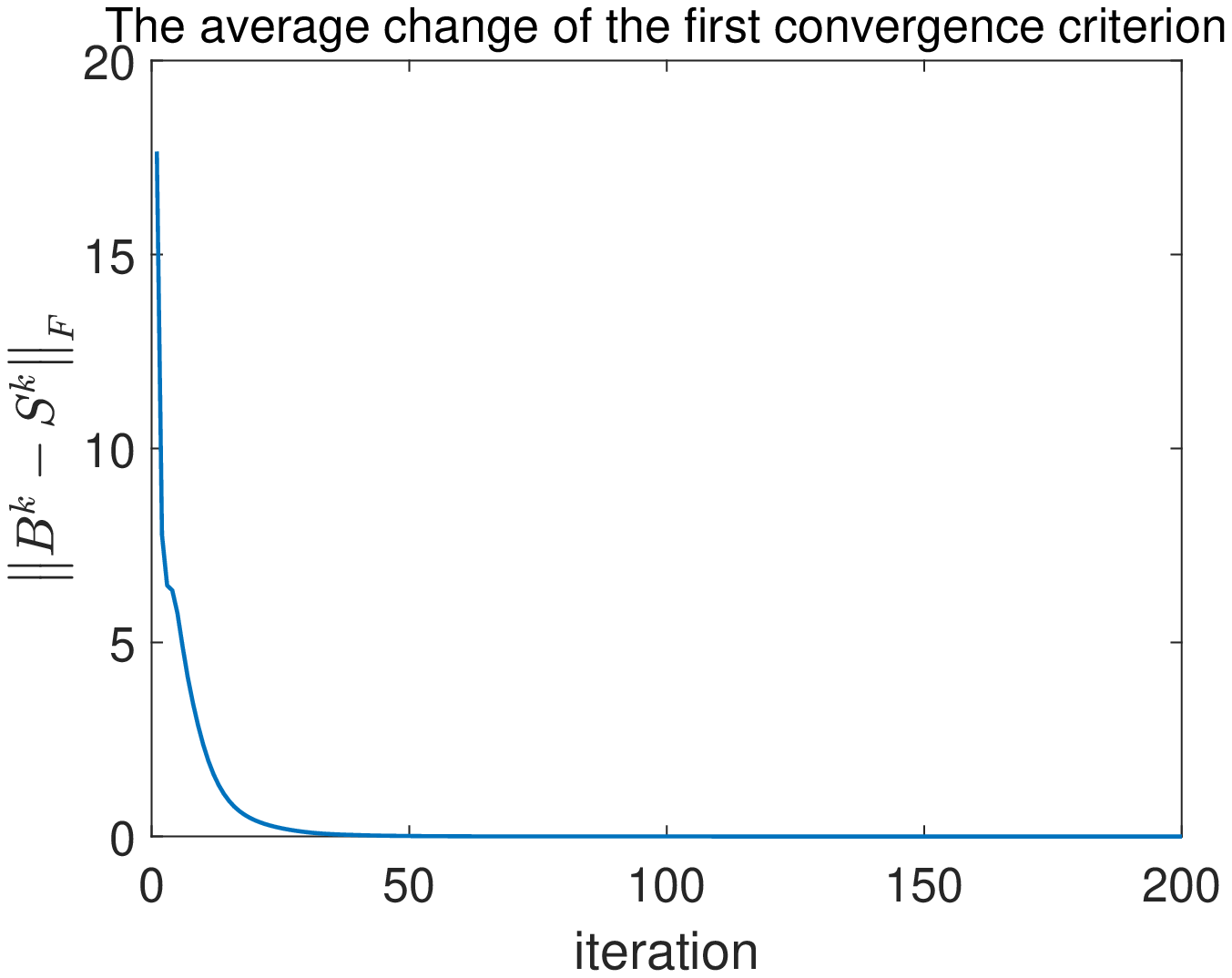}}
	\end{minipage}
	\begin{minipage}{0.32\linewidth}
		\vspace{3pt}
		\centerline{\includegraphics[width=\textwidth]{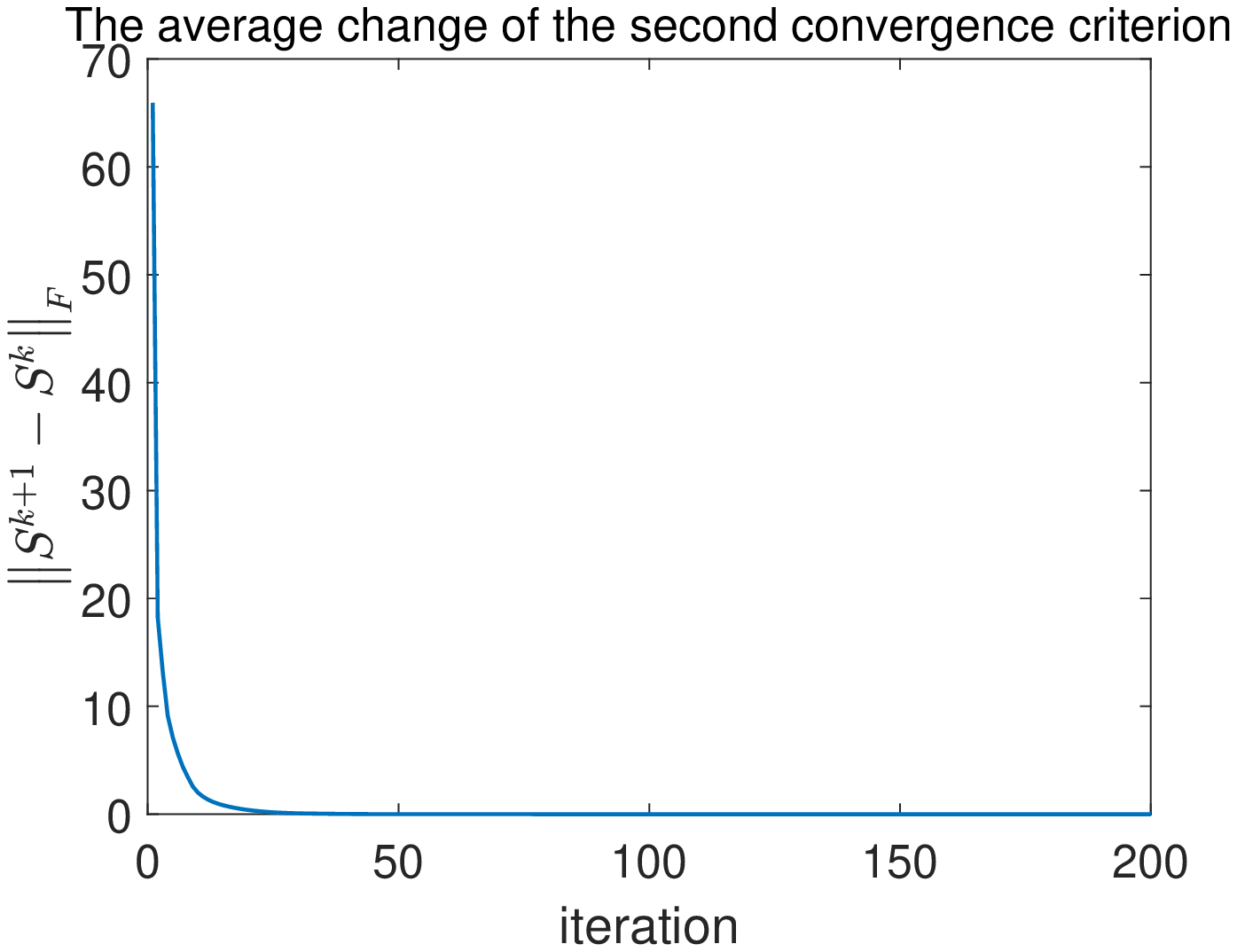}}
	\end{minipage}
	\begin{minipage}{0.32\linewidth}
		\vspace{3pt}
		\centerline{\includegraphics[width=\textwidth]{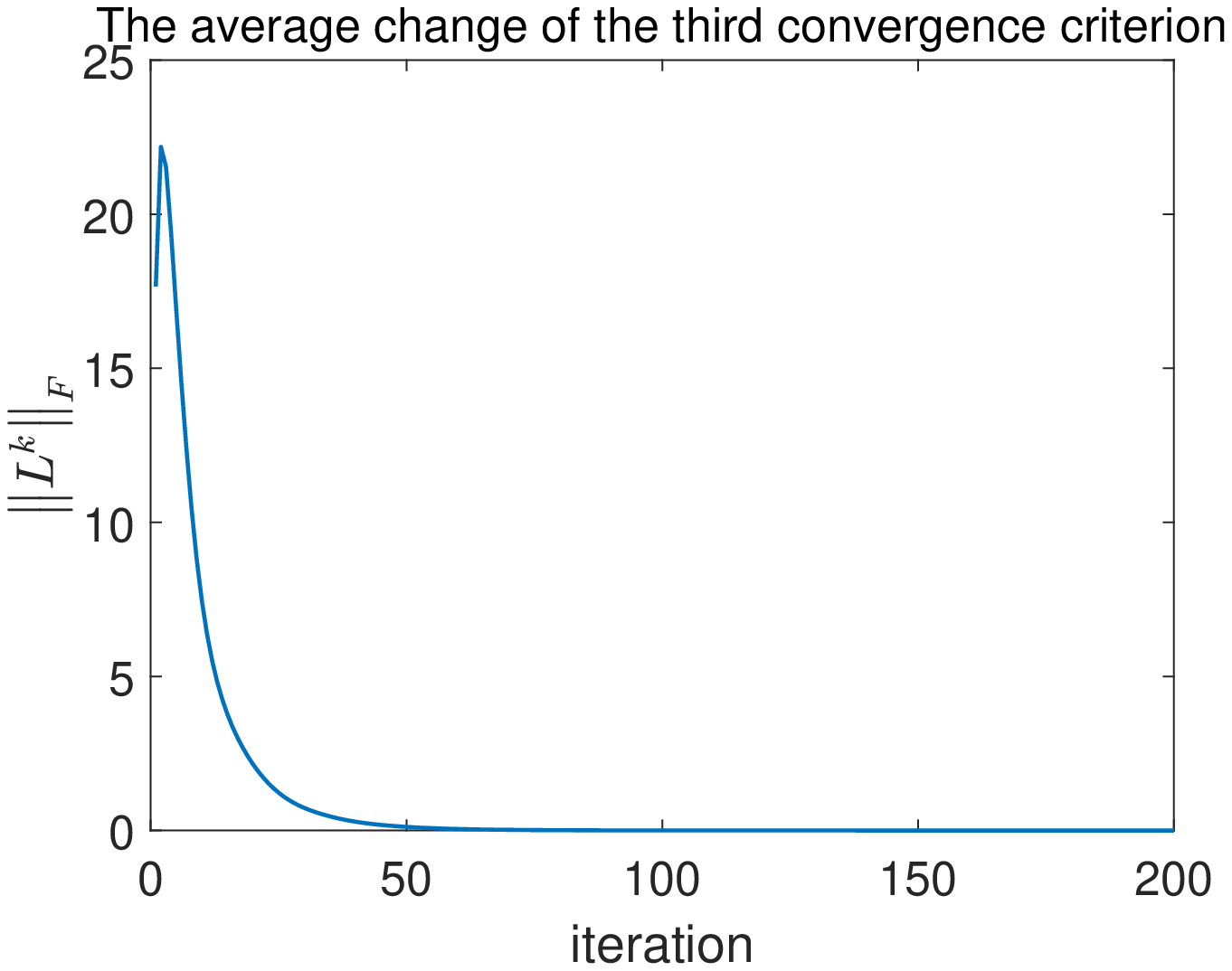}}
	\end{minipage}
	\caption{The average change of convergence criterion in iterations over 10 experiments ($N=500,M=150,K=50,J=10$)}
	\label{fig:1}
\end{figure}

In the second experiment, we tested the convergence criterion (\ref{equ:4.7}) given by Theorem \ref{thm:5.3}. We set $N=500,M=150,K=50,J=10$. The aim was to observe the average change in the convergence criterion (\ref{equ:4.7}) in iterations over 10 random repetitions. The average change of the convergence criterion in 200 iterations is shown in Figure \ref{fig:1}. All three variables in (\ref{equ:4.7}) tend to zero and reach an order of magnitude of $-16$ at $MaxIter=10^{3}$; this satisfies our analysis in Theorem \ref{thm:5.3}.
 
Then, we studied the influence of the convergence criterion (\ref{equ:4.7}) on our algorithm. When all three variables in the convergence criterion (\ref{equ:4.7}) were less than $10^{-6}$, the proposed algorithm was considered to have found an approximate optimal solution, and the iteration was stopped. As shown in Table \ref{tab:3}, with the convergence criterion, the proposed algorithm can successfully recover the sparse signals in a short time.

In the following experiments, we applied MMV-ADMM-$\ell_{2,0}$ with the proposed convergence criterion.

\begin{table}[hbt!]
    \centering
    \begin{tabular}{ccccc}
 \hline
\multicolumn{1}{l}{data} & \multicolumn{2}{l}{Without convergence criterion} & \multicolumn{2}{l}{With convergence criterion} \\ 
           & RMSE & Time(s) & RMSE & Time(s) \\ \hline
         1 & 8.2155e-16 & 0.5765 & 9.4005e-10 & 0.1560 \\ 
         2 & 9.0799e-16 & 0.5193 & 7.4102e-10 & 0.1179  \\ 
         3 & 7.6515e-16 & 0.4940 & 6.9234e-10 & 0.1116  \\ 
         4 & 8.4516e-16 & 0.5109 & 8.8110e-10 & 0.1132  \\ 
         5 & 8.5083e-16 & 0.5278 & 7.4768e-10 & 0.1269  \\ 
         6 & 9.4163e-16 & 0.5697 & 8.0561e-10 & 0.1508  \\ 
         7 & 8.2227e-16 & 0.5152 & 6.5987e-10 & 0.1108  \\ 
         8 & 8.9286e-16 & 0.5053 & 7.4714e-10 & 0.1069  \\ 
         9 & 7.6474e-16 & 0.4863 & 8.1573e-10 & 0.0949  \\ 
         10 & 8.0768e-16 & 0.4796 & 7.8128e-10 & 0.1024  \\ \hline
         Mean & 8.4199e-16 & 0.5185 & 7.8118e-10 & 0.1191  \\ 
         Std & 5.8515e-17 & 0.0324 & 8.4008e-11 & 0.0200   \\ \hline
    \end{tabular}
    \caption{The average recovery quality over 10 random experiments with and without convergence criterion($N=500,M=150,K=50,J=10$)}
    \label{tab:3}
\end{table}

\subsection{Performance with different sparsity $K$}\label{sub:6.4}

\begin{figure}[hbt!]
	
	\begin{minipage}{0.48\linewidth}
		\vspace{3pt}
		\centerline{\includegraphics[width=\textwidth]{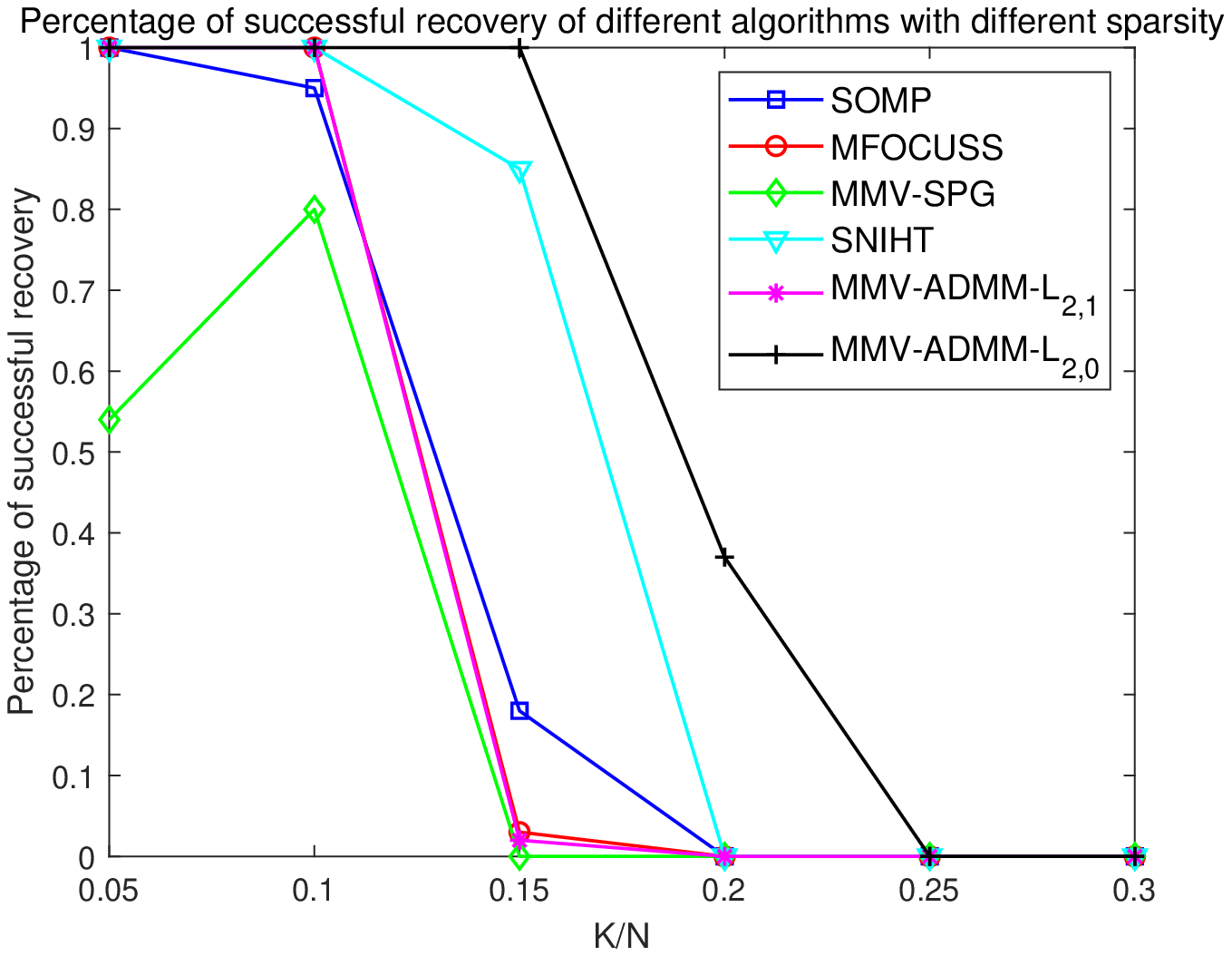}}
	\end{minipage}
	\begin{minipage}{0.48\linewidth}
		\vspace{3pt}
		\centerline{\includegraphics[width=\textwidth]{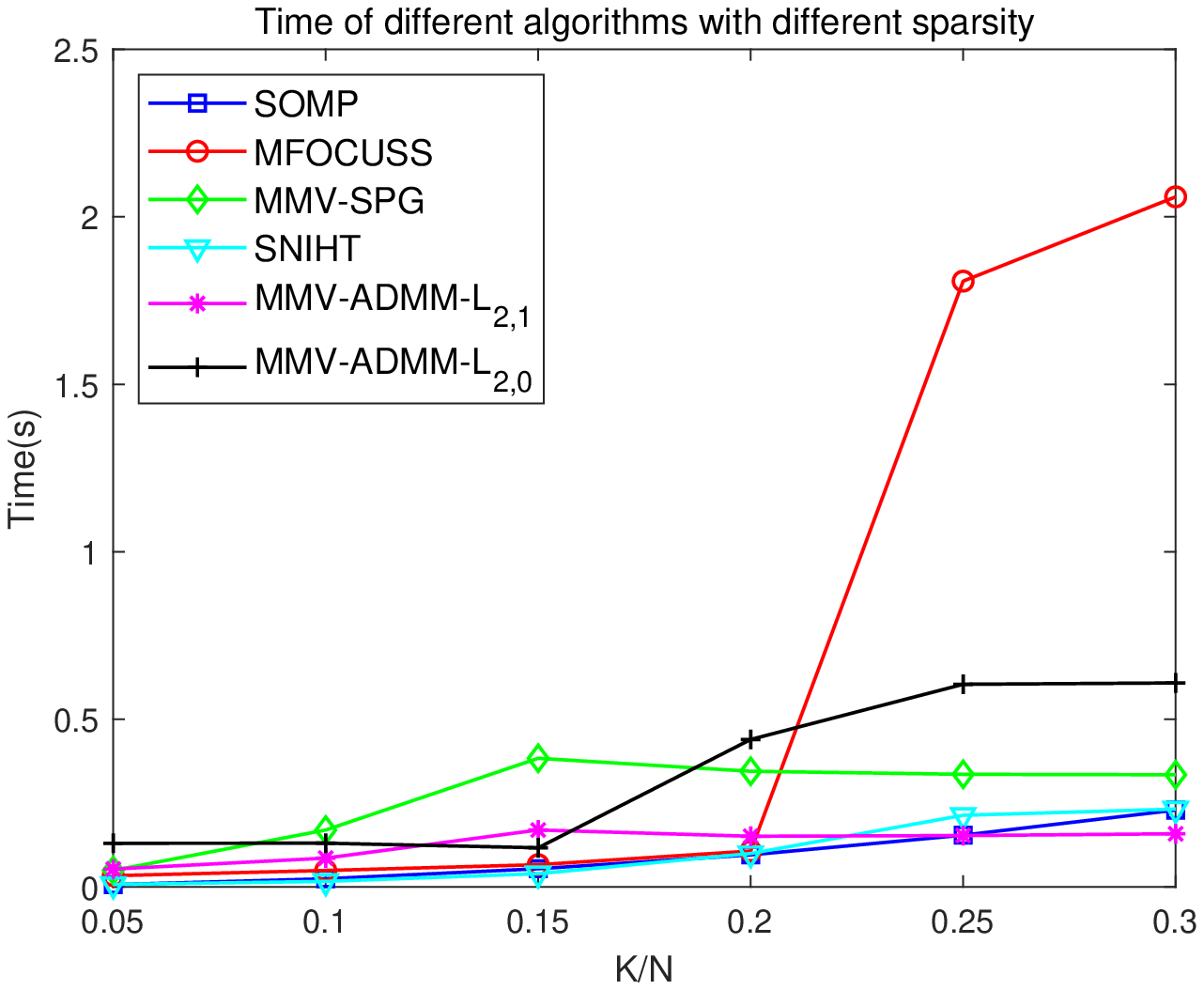}}
	\end{minipage}
	\caption{The recovery quality of different algorithms with different sparsity $K$ ($N=500,M=150,J=10$)}
	\label{fig:2}
\end{figure}

In the third experiment, we studied how sparsity influenced the recovery quality of all these algorithms. Set $N=500,M=150,J=10$ and let $K$ range from 25 to 150 with step size 25. Observe how the percentage of successful recovery changes with $\frac{K}{N}$ when applying different algorithms. The experiment results are shown in Figure \ref{fig:2}.

It is not hard to see that all these algorithms fail to recover the original signals when the sparsity of signals is large ($i.e.$, $\frac{K}{N}\ge 0.2$). However, when the sparsity is not so large, that is, $0.1\le \frac{K}{N}\le 0.15$, the proposed algorithm can successfully recover all signals over 100 experiments, whereas the other algorithms have poor performance. In addition, in the case of successful recovery, all these algorithms are fast enough. Therefore, we can conclude that for the MMV problem, when the original signals are not sparse enough, the proposed algorithm is more suitable.

\subsection{Performance with different undersampling rate $\frac{N}{M}$}\label{sub:6.5}

The main purpose of compressed sensing is to reduce the number of measurements while ensuring good recovery quality. To illustrate that the proposed algorithm has better performance with a larger undersampling rate compared with that of other algorithms, we studied the percentage of successful recovery with different undersampling rates. We set $N=500,K=50,J=10$ and let the undersampling rate $\frac{N}{M}$ range from 1.6 to 8 with a step size of 1.6.

\begin{figure}[hbt!]
	
	\begin{minipage}{0.48\linewidth}
		\vspace{3pt}
		\centerline{\includegraphics[width=\textwidth]{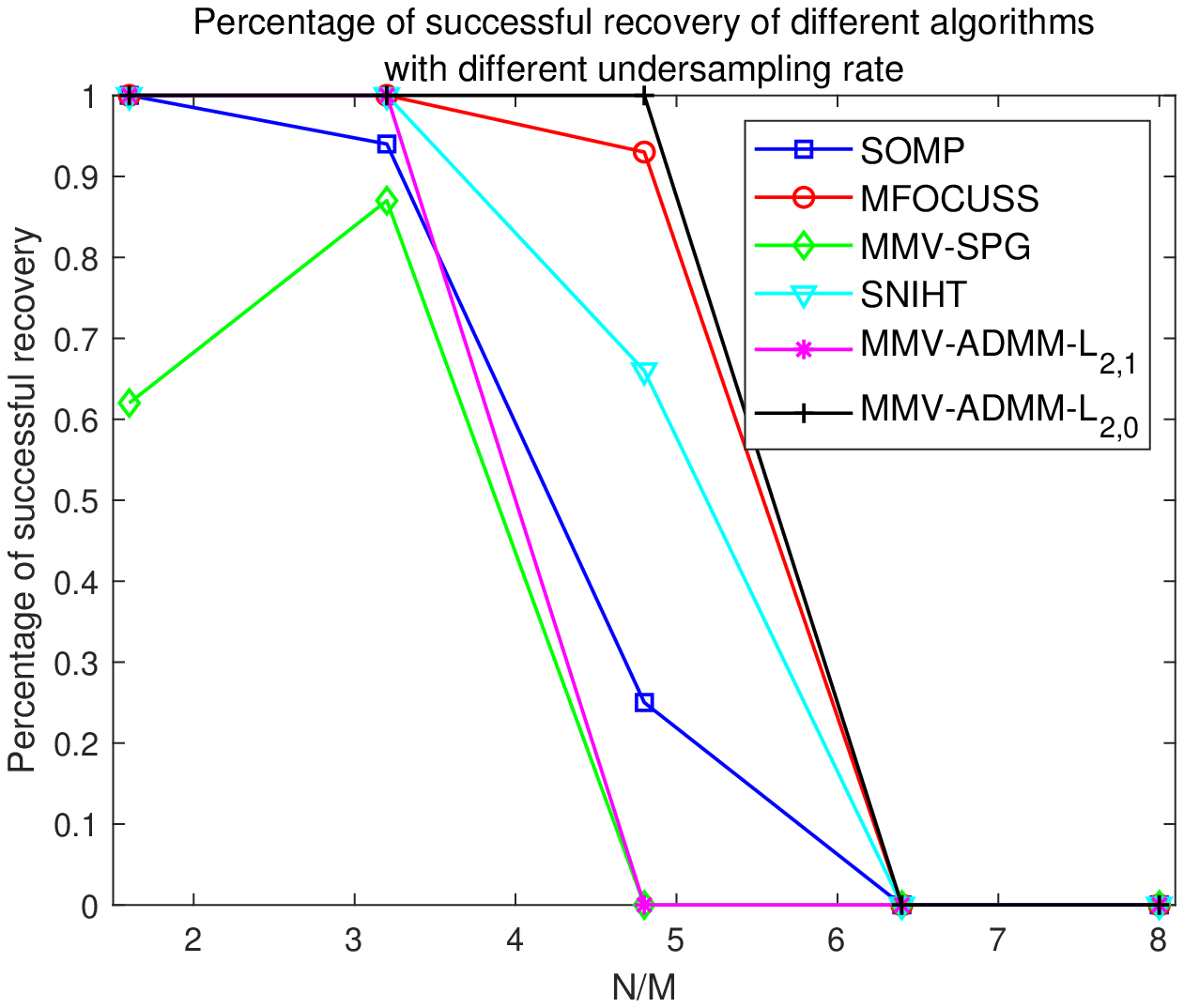}}
	\end{minipage}
	\begin{minipage}{0.48\linewidth}
		\vspace{3pt}
		\centerline{\includegraphics[width=\textwidth]{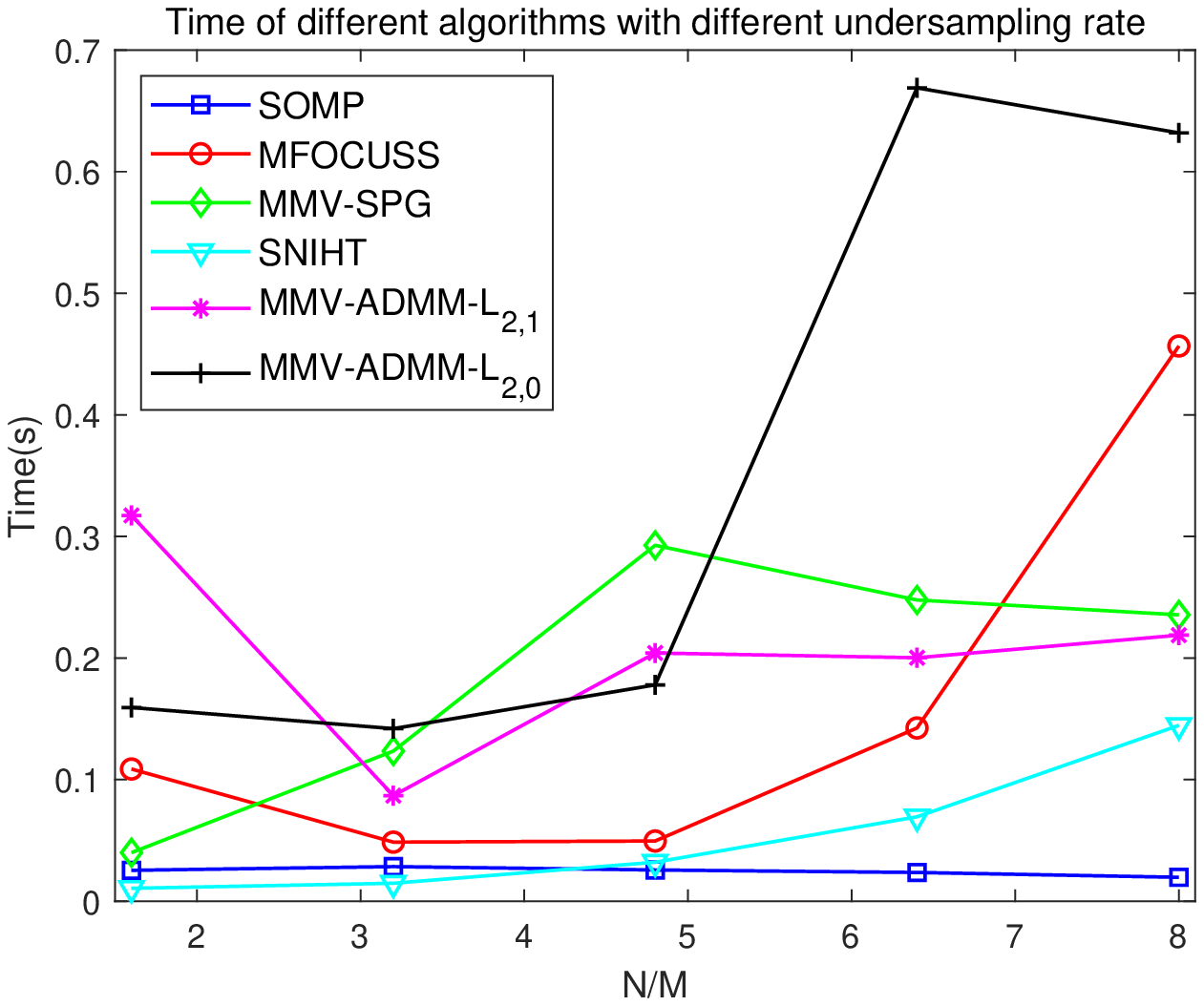}}
	\end{minipage}
	\caption{The recovery quality of different algorithms with different undersampling rate $\frac{N}{M}$ ($N=500,K=50,J=10$)}
	\label{fig:3}
\end{figure}

As shown in Figure \ref{fig:3}, when the undersampling rate $\frac{N}{M}\ge 6.4$, all algorithms fail to recover the original signals. However, when $\frac{N}{M}= 4.8$, the proposed algorithm can still recover all signals, whereas MFOCUSS can only recover approximately 94\% of the signals and other algorithms perform poorly. With regard to the running time, we just need to consider the case of successful recovery. We see that the proposed algorithm has average performance among all algorithms. This implies that the proposed algorithm has a larger undersampling rate compared with that of the other algorithms; this is the key of compressed sensing.

\subsection{Performance with different number of sensors $J$}\label{sub:6.6}

In this experiment, we studied how the recovery quality was affected by the number of sensors $J$. We set $N=500,M=150,K=50$ and let $J$ range from 1 (multiplied by 2 per step) to 32.

\begin{figure}[hbt!]
	
	\begin{minipage}{0.48\linewidth}
		\vspace{3pt}
		\centerline{\includegraphics[width=\textwidth]{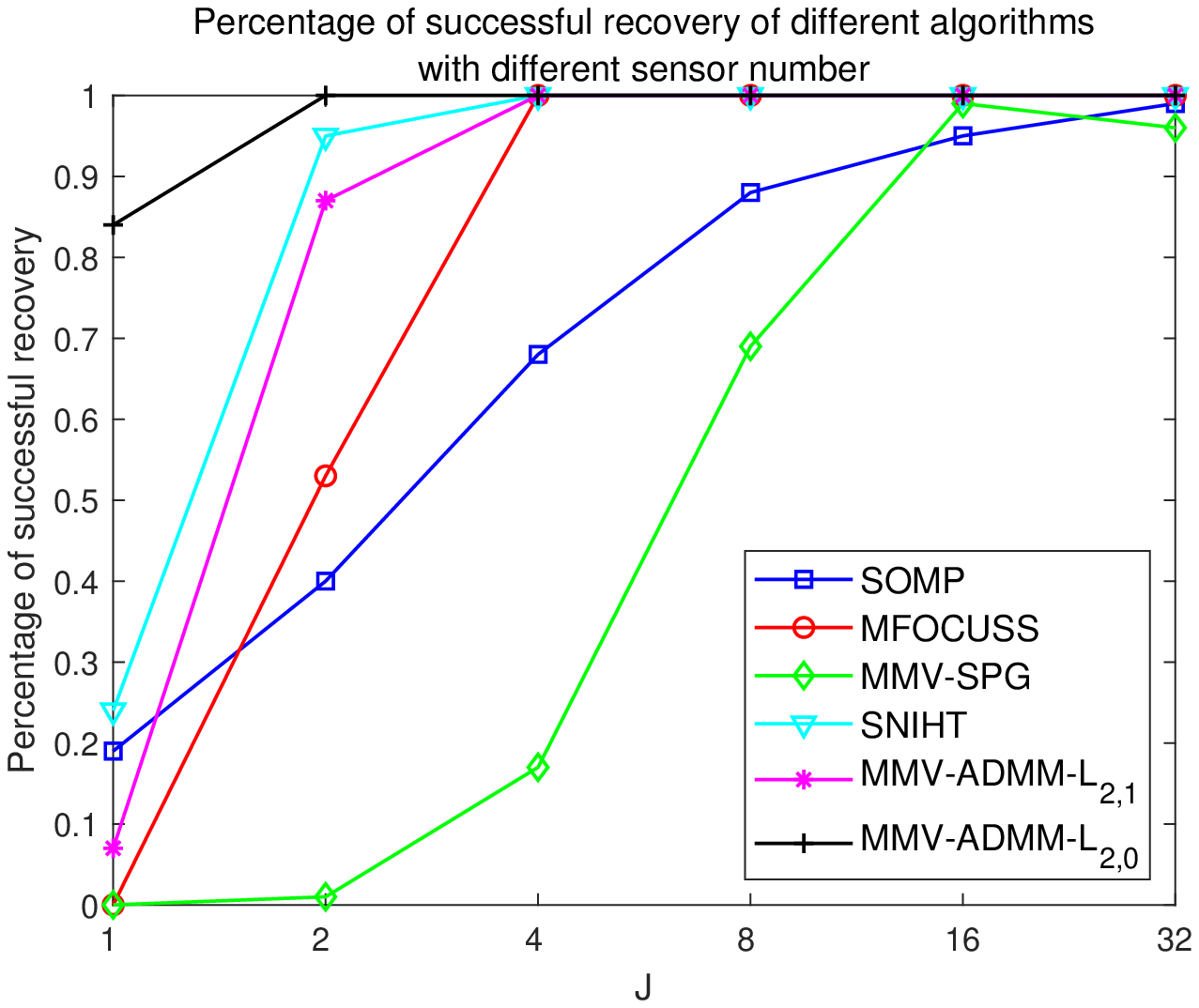}}
	\end{minipage}
	\begin{minipage}{0.48\linewidth}
		\vspace{3pt}
		\centerline{\includegraphics[width=\textwidth]{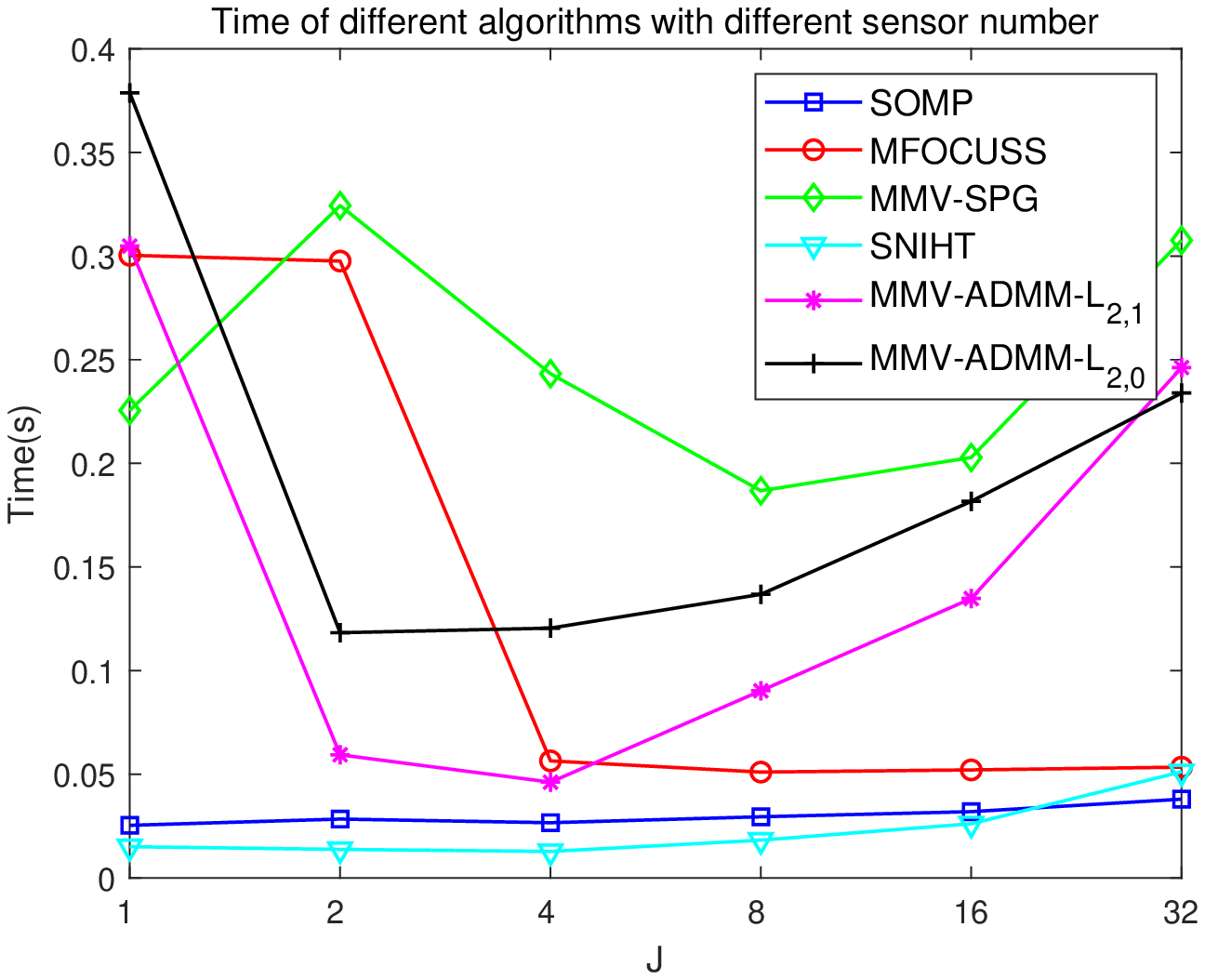}}
	\end{minipage}
	\caption{The recovery quality of different algorithms with different number of sensors $J$ ($N=500,M=150,K=50$)}
	\label{fig:4}
\end{figure}

Figure \ref{fig:4} shows that when the number of sensors is 1 ($i.e.$, a SMV problem), all algorithms cannot successfully recover all signals. However, the proposed algorithm can recover approximately 82\% of signals, whereas the other algorithms perform poorly. When $J\ge2$, the proposed algorithm can successfully recover all signals. However, the other algorithms can only recover all signals when $J\ge 4$. The running time of the proposed algorithm is the average of that of all the other algorithms, all of which are fast enough to recover the original signals. This experiment proves that the proposed algorithm performs well even though with a small number of sensors, whereas the other algorithms cannot.

\subsection{Performance with different sample dimensions $N$}\label{sub:6.7}

We tested the recovery quality of our algorithm for varying numbers of dimensions. We set $J=10$ and tested $N=100,500,1000,1500,3000$ with $\frac{N}{M}=3$ and $\frac{K}{N}=0.1$.

\begin{figure}[hbt!]
	
	\begin{minipage}{0.48\linewidth}
		\vspace{3pt}
		\centerline{\includegraphics[width=\textwidth]{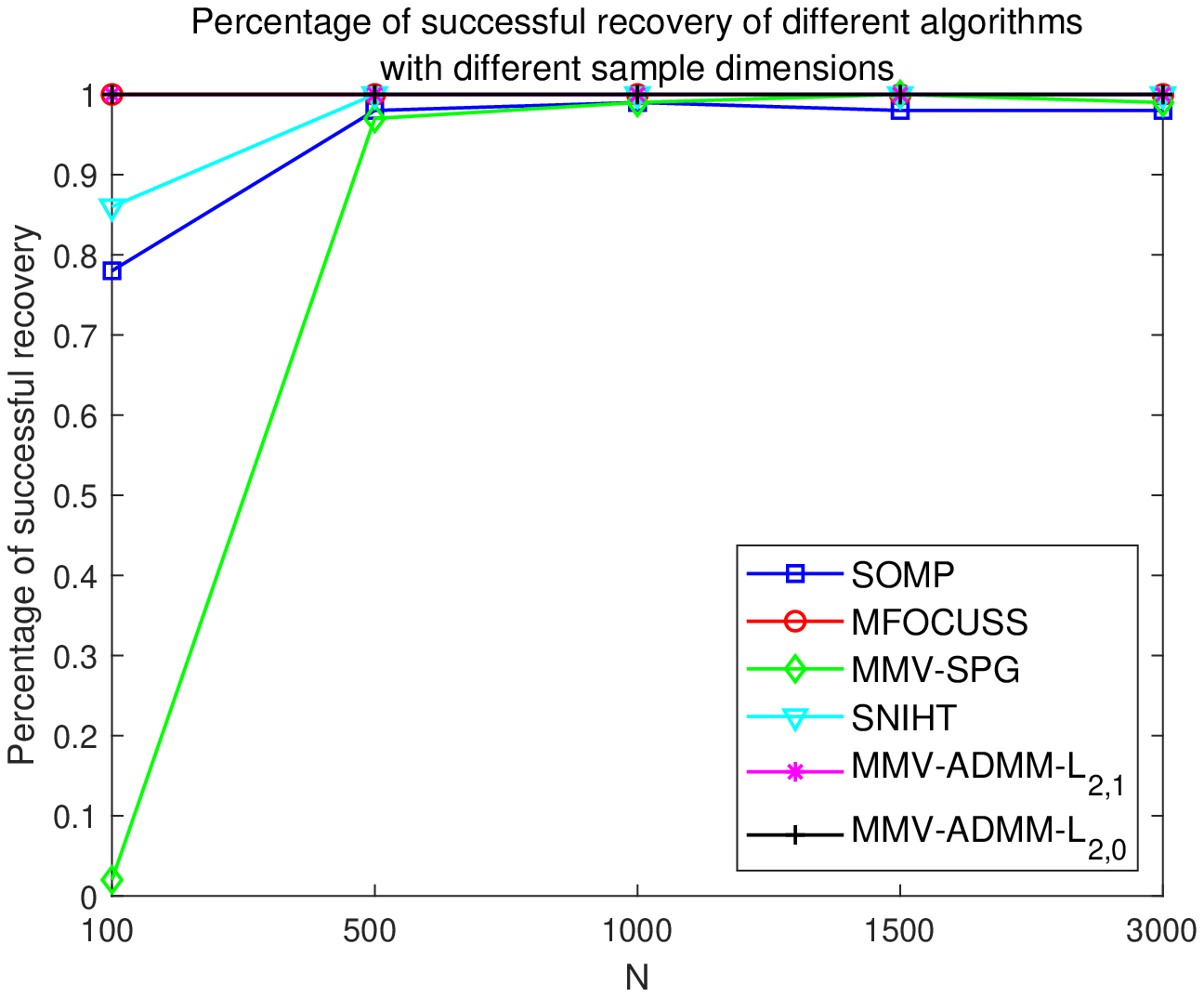}}
	\end{minipage}
	\begin{minipage}{0.48\linewidth}
		\vspace{3pt}
		\centerline{\includegraphics[width=\textwidth]{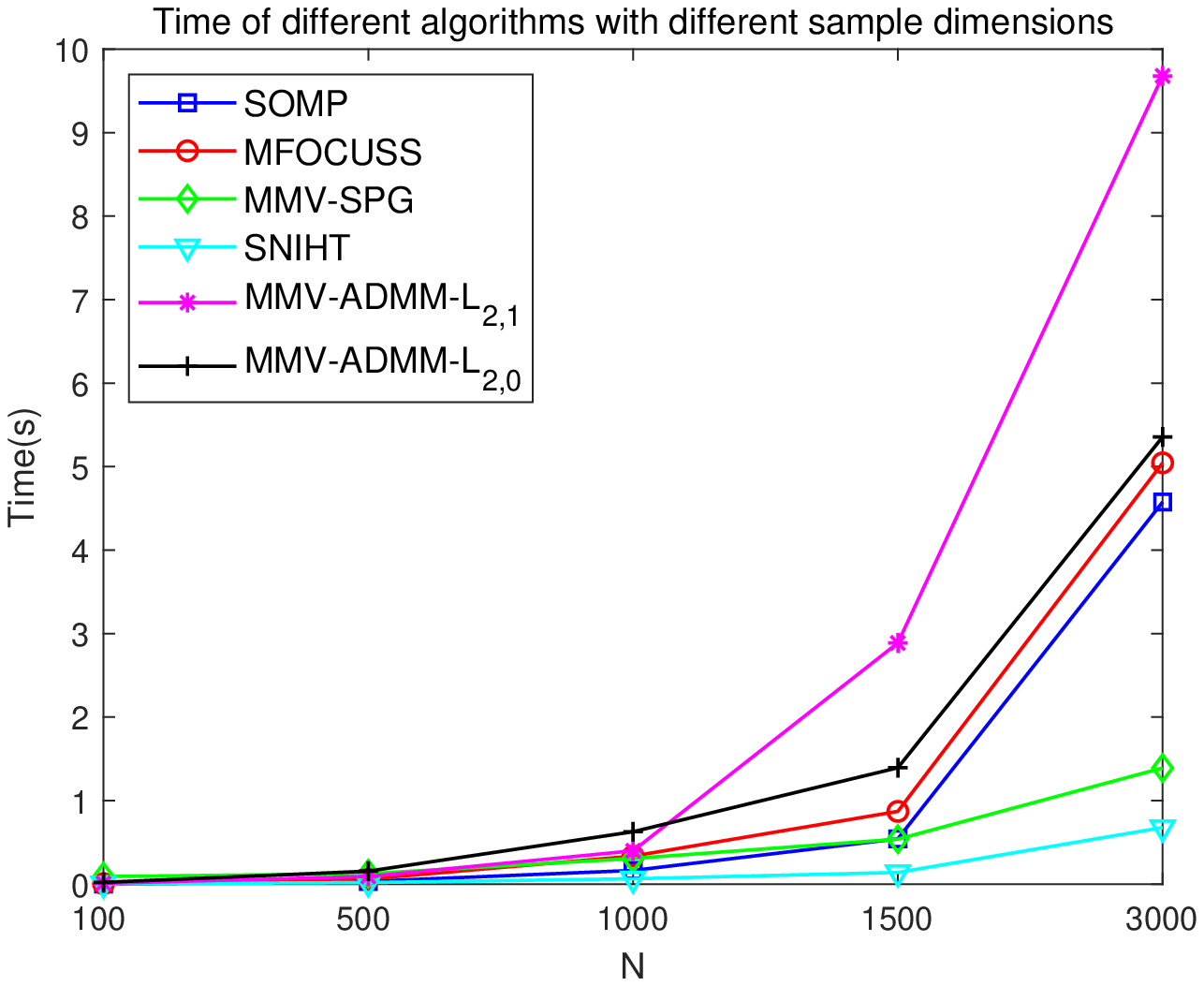}}
	\end{minipage}
	\caption{The recovery quality of different algorithms with different sample dimensions $N$ ($\frac{N}{M}=3,\frac{K}{N}=0.1,J=10$)}
	\label{fig:5}
\end{figure}

Figure \ref{fig:5} shows that MFOCUSS, MMV-ADMM-$\ell_{2,1}$, and the proposed algorithm all perform well irrespective of how the dimension $N$ changes. MFOCUSS and the proposed algorithm have similar running times; however, the running time of MMV-ADMM-$\ell_{2,1}$ is higher when $N$ is large. This experiment proves that the proposed algorithm is also suitable for a large number of dimensions.

\subsection{Performance with SMW-formula}\label{sub:6.8}

Finally, we designed an experiment to illustrate the advantage of MMV-ADMM-$\ell_{2,0}$ with the SMW formula when $N$ is large. We set $N=5000,M=1500,K=500,J=10$ and observe the RMSE and running time over 10 random experiments.

As shown in Table \ref{tab:4}, with the SMW-formula, MMV-ADMM-$\ell_{2,0}$-SMW requires less time to obtain solutions with the same precision as that of MMV-ADMM-$\ell_{2,0}$. 

\begin{table}[hbt!]
    \centering
    \begin{tabular}{ccccc}
 \hline
\multicolumn{1}{l}{data} & \multicolumn{2}{l}{MMV-ADMM-$\ell_{2,0}$} & \multicolumn{2}{l}{MMV-ADMM-$\ell_{2,0}$-SMW} \\ 
           & RMSE & Time(s) & RMSE & Time(s) \\ \hline
         1 & 2.5689e-10 & 12.5324 & 2.3941e-10 & 11.5070 \\ 
         2 & 2.6148e-10 & 13.4957 & 2.4949e-10 & 12.0098  \\ 
         3 & 2.5017e-10 & 13.3999 & 2.6850e-10 & 12.7097  \\ 
         4 & 2.5103e-10 & 13.8419 & 2.5707e-10 & 12.2158  \\ 
         5 & 2.4778e-10 & 13.6656 & 2.4574e-10 & 12.1301  \\ 
         6 & 2.4846e-10 & 13.5132 & 2.5370e-10 & 12.0958  \\ 
         7 & 2.4829e-10 & 13.5649 & 2.4438e-10 & 12.4199  \\ 
         8 & 2.5644e-10 & 14.2626 & 2.5308e-10 & 12.4762  \\ 
         9 & 2.6028e-10 & 13.6527 & 2.5655e-10 & 12.3948  \\ 
         10 & 2.4507e-16 & 13.8025 & 2.5998e-10 & 12.3597  \\ \hline
         Mean & 2.5259e-10 & 13.5732 & 2.5279e-10 & 12.2319  \\ 
         Std & 5.7252e-12 & 0.4394 & 8.4480e-12 & 0.3283   \\ \hline
    \end{tabular}
    \caption{The average recovery quality over 10 random experiments with and without SMW-formula($N=5000,M=1500,K=500,J=10$)}
    \label{tab:4}
\end{table}

\section{Conclusions}\label{sec:7}

In this study, we propose MMV-ADMM-$\ell_{2,0}$, an ADMM algorithm based on $\ell_{2,0}$-norm for the MMV problem. Instead of using the relaxation $\ell_{2,1}$-norm, we directly solve the $\ell_{2,0}$-norm minimization problem; this is the main difference between our study and other works \cite{16,7,20}. We prove that the proposed algorithm has global convergence and determine its convergence criterion (\ref{equ:4.7}). Through numerical simulations, we test the validity of the proposed algorithm and its convergence criterion. The experiment results are consistent with Theorems \ref{thm:5.2} and \ref{thm:5.3}. In comparisons against other algorithms, MMV-ADMM-$\ell_{2,0}$ shows better performance with high sparsity $K$ and few sensors $J$, and it is insensitive to the dimension $N$. Moreover, MMV-ADMM-$\ell_{2,0}$ has a larger undersampling rate compared with those of other algorithms, especially MMV-ADMM-$\ell_{2,1}$, which is the key of compressed sensing.

The proposed algorithm has some drawbacks. Although its running time is roughly the average of those of the other algorithms, its efficiency remains lower than those of SOMP and SNIHT. We propose MMV-ADMM-$\ell_{2,0}$-SMW to reduce the computational cost, and it runs faster. However, this improvement is not significant until the dimension $N$ becomes large.

The research in this paper can still be studied through a deep dive. First, Algorithm \ref{alg:1} can be extended to the generalised JSM-2 problem, where multiple vectors are measured by different sensing matrices. And the derivation of the algorithm for the generalized JSM-2 problem is nearly the same as discussions in Section \ref{sec:3} and \ref{sec:4} of this paper. Second, Algorithm \ref{alg:1} can accelerate even further. The most time cost step in the iterations of Algorithm \ref{alg:1} is updating $S$ by (\ref{equ:4.5}). The key is to solve a high-dimensional system of positive-definite linear equations. Drawing on recent advances in the field of numerical algebra, $S$ can be updated more efficiently, thus Algorithm \ref{alg:1} can converge faster. We will focus on the above two points in the future research; give a more efficient algorithm for the MMV problem, and extend it to the generalized JSM-2 problem.

\bibliographystyle{unsrt}
\bibliography{references}

\begin{thebibliography}{10}

\bibitem{1}
Ewout van~den Berg and Michael~P. Friedlander.
\newblock Theoretical and empirical results for recovery from multiple
  measurements.
\newblock {\em IEEE Transactions on Information Theory}, 56(5):2516--2527,
  2010.

\bibitem{11}
D.L. Donoho.
\newblock Compressed sensing.
\newblock {\em IEEE Transactions on Information Theory}, 52(4):1289--1306,
  2006.

\bibitem{12}
E.J. Candes, J.~Romberg, and T.~Tao.
\newblock Robust uncertainty principles: exact signal reconstruction from
  highly incomplete frequency information.
\newblock {\em IEEE Transactions on Information Theory}, 52(2):489--509, 2006.

\bibitem{13}
David~L. Donoho and Michael Elad.
\newblock Optimally sparse representation in general (nonorthogonal)
  dictionaries via $\ell_{1}$ minimization.
\newblock {\em Proceedings of the National Academy of Sciences of the United
  States of America}, 100:2197 -- 2202, 2003.

\bibitem{14}
Scott~Saobing Chen, David~L. Donoho, and Michael~A. Saunders.
\newblock Atomic decomposition by basis pursuit.
\newblock {\em SIAM J. Sci. Comput.}, 20:33--61, 1998.

\bibitem{15}
Qing Qu, Nasser~M. Nasrabadi, and Trac~D. Tran.
\newblock Abundance estimation for bilinear mixture models via joint sparse and
  low-rank representation.
\newblock {\em IEEE Transactions on Geoscience and Remote Sensing},
  52(7):4404--4423, 2014.

\bibitem{16}
Shane~F. Cotter, Bhaskar~D. Rao, Kjersti Engan, and Kenneth Kreutz-Delgado.
\newblock Sparse solutions to linear inverse problems with multiple measurement
  vectors.
\newblock {\em IEEE Transactions on Signal Processing}, 53:2477--2488, 2005.

\bibitem{17}
Irina~F. Gorodnitsky, John~S. George, and B.D. Rao.
\newblock Neuromagnetic source imaging with focuss: a recursive weighted
  minimum norm algorithm.
\newblock {\em Electroencephalography and clinical neurophysiology}, 95
  4:231--51, 1995.

\bibitem{18}
Dmitry~M. Malioutov, M{\"u}jdat Çetin, and Alan~S. Willsky.
\newblock Source localization by enforcing sparsity through a laplacian prior:
  an svd-based approach.
\newblock {\em IEEE Workshop on Statistical Signal Processing, 2003}, pages
  573--576, 2004.

\bibitem{19}
S.F. Cotter and B.D. Rao.
\newblock Sparse channel estimation via matching pursuit with application to
  equalization.
\newblock {\em IEEE Transactions on Communications}, 50(3):374--377, 2002.

\bibitem{7}
Jie Chen and Xiaoming Huo.
\newblock Theoretical results on sparse representations of multiple-measurement
  vectors.
\newblock {\em IEEE Transactions on Signal Processing}, 54(12):4634--4643,
  2006.

\bibitem{20}
Yonina~C. Eldar and Moshe Mishali.
\newblock Robust recovery of signals from a structured union of subspaces.
\newblock {\em IEEE Transactions on Information Theory}, 55(11):5302--5316,
  2009.

\bibitem{21}
Joel~A. Tropp, Anna~C. Gilbert, and Martin Strauss.
\newblock Algorithms for simultaneous sparse approximation. part i: Greedy
  pursuit.
\newblock {\em Signal Process.}, 86:572--588, 2006.

\bibitem{28}
Jeffrey~D. Blanchard, Michael Cermak, David Hanle, and Yirong Jing.
\newblock Greedy algorithms for joint sparse recovery.
\newblock {\em IEEE Transactions on Signal Processing}, 62(7):1694--1704, 2014.

\bibitem{8}
Stephen~P. Boyd, Neal Parikh, Eric~King wah Chu, Borja Peleato, and Jonathan
  Eckstein.
\newblock Distributed optimization and statistical learning via the alternating
  direction method of multipliers.
\newblock {\em Found. Trends Mach. Learn.}, 3:1--122, 2011.

\bibitem{4}
Wen Zaiwen, Hu~Jiang, Li~Yongfeng, and Liu Haoyang.
\newblock {\em Optimization: Modeling, Algorithm and Theory(in Chinese)}.
\newblock Higher Education Press, 2020.

\bibitem{3}
Stephen~P. Boyd and Lieven Vandenberghe.
\newblock Convex optimization.
\newblock {\em IEEE Transactions on Automatic Control}, 51:1859--1859, 2004.

\bibitem{5}
Attouch H, Bolte J, and Svaiter B.~F.
\newblock Convergence of descent methods for semi-algebraic and tame problems:
  proximal algorithms, forward-backward splitting, and regularized gauss-seidel
  methods.
\newblock {\em Mathematical Programming}, 2013.

\bibitem{2}
Boris~S. Mordukhovich.
\newblock {\em Variational analysis and generalized differentiation I: Basic
  theory}.
\newblock Grundlehren der mathematischen Wissenschaften. Springer Berlin,
  Heidelberg, 2006.

\bibitem{6}
Bolte J, Sabach S, and Teboulle M.
\newblock Proximal alternating linearized minimization for nonconvex and
  nonsmooth problems.
\newblock {\em Math. Program. 146, 459–494}, 2014.

\bibitem{27}
Simon Foucart.
\newblock Hard thresholding pursuit: An algorithm for compressive sensing.
\newblock {\em SIAM J. Numer. Anal.}, 49:2543--2563, 2011.

\bibitem{24}
Jack Sherman and Winifred~J. Morrison.
\newblock Adjustment of an inverse matrix corresponding to a change in one
  element of a given matrix.
\newblock {\em Annals of Mathematical Statistics}, 21:124--127, 1950.

\bibitem{25}
M~Woodbury.
\newblock {\em Inverting Modified Matrices}.
\newblock Statistical Research Group, Princeton University, Princeton, 1950.

\bibitem{9}
Ke~Guo, Deren Han, and Tingting Wu.
\newblock Convergence of alternating direction method for minimizing sum of two
  nonconvex functions with linear constraints.
\newblock {\em International Journal of Computer Mathematics}, 94:1653 -- 1669,
  2017.

\bibitem{10}
Oleg~P. Burdakov, Christian Kanzow, and Alexandra Schwartz.
\newblock Mathematical programs with cardinality constraints: Reformulation by
  complementarity-type conditions and a regularization method.
\newblock {\em SIAM J. on Optimization}, 26(1):397–425, jan 2016.

\bibitem{23}
Ewout van~den Berg and Michael~P. Friedlander.
\newblock Sparse optimization with least-squares constraints.
\newblock {\em SIAM J. Optim.}, 21:1201--1229, 2011.

\end{thebibliography}

\end{document}